\documentclass[useAMS,usenatbib]{mn2e}
\usepackage{graphicx}
 \usepackage{aas_macros}
\usepackage{etoolbox}
\usepackage{multirow}
\makeatletter

\patchcmd{\NAT@citex}
  {\@citea\NAT@hyper@{%
     \NAT@nmfmt{\NAT@nm}%
     \hyper@natlinkbreak{\NAT@aysep\NAT@spacechar}{\@citeb\@extra@b@citeb}%
     \NAT@date}}
  {\@citea\NAT@nmfmt{\NAT@nm}%
   \NAT@aysep\NAT@spacechar\NAT@hyper@{\NAT@date}}{}{}

\patchcmd{\NAT@citex}
  {\@citea\NAT@hyper@{%
     \NAT@nmfmt{\NAT@nm}%
     \hyper@natlinkbreak{\NAT@spacechar\NAT@@open\if*#1*\else#1\NAT@spacechar\fi}%
       {\@citeb\@extra@b@citeb}%
     \NAT@date}}
  {\@citea\NAT@nmfmt{\NAT@nm}%
   \NAT@spacechar\NAT@@open\if*#1*\else#1\NAT@spacechar\fi\NAT@hyper@{\NAT@date}}
  {}{}

\makeatother

\title[S999 mass ratio]{How elevated is the dynamical-to-stellar mass ratio of the ultra-compact dwarf S999?}
\author[Janz et al.]{Joachim Janz$^{1}$\thanks{E-mail:
jjanz@swin.edu.au}, Duncan A. Forbes$^{1}$, Mark A. Norris$^{2}$,  Jay Strader$^3$, \newauthor  Samantha J. Penny$^{4,1,5}$, Martina Fagioli$^{1,6}$, Aaron J. Romanowsky$^{7,8}$ \\
$^{1}$Centre for Astrophysics \& Supercomputing, Swinburne University, Hawthorn, VIC 3122, Australia\\
$^{2}$Max Planck Institut f\"ur Astronomie, K\"onigstuhl 17, D-69117, Heidelberg, Germany\\
$^3$Department of Physics and Astronomy, Michigan State University, East Lansing, Michigan 48824, USA\\
$^{4}$School of Physics, Monash University, Clayton, Victoria 3800, Australia\\
$^{5}$Institute of Cosmology and Gravitation, University of Portsmouth, Dennis Sciama Building, Burnaby Road, Portsmouth,\\ PO1 3FX, United Kingdom\\
$^6$Institute for Astronomy, ETH Z\"urich, Wolfgang-Pauli-Str.\ 27, 8093 Z\"urich, Switzerland\\
$^{7}$Department of Physics and Astronomy, San Jos\'e State University, San Jose, CA 95192, USA\\
$^{8}$University of California Observatories, 1156 High Street, Santa Cruz, CA 95064, USA}
\begin{document}

\date{Accepted 2014 --. Received 2014 --; in original form 2014 --}

\pagerange{\pageref{firstpage}--\pageref{lastpage}} \pubyear{2014}

\maketitle

\label{firstpage}

\begin{abstract}
Here we present new Keck ESI high-resolution spectroscopy and deep archival {\it HST}/ACS imaging for S999, an ultra-compact dwarf in the vicinity of M87, which was claimed to have an extremely high  dynamical-to-stellar mass ratio. Our data increase the total integration times by a factor of 5 and  60 for spectroscopy and imaging, respectively. This allows us to constrain  the stellar population parameters for the first time (simple stellar population equivalent age $=7.6^{+2.0}_{-1.6}$ Gyr; $[Z/\textrm{H}]=-0.95^{+0.12}_{-0.10}$;  $[\alpha/\textrm{Fe}]=0.34^{+0.10}_{-0.12}$). 
Assuming a Kroupa stellar initial mass function, the stellar population parameters and luminosity ($M_{F814W}=-12.13\pm0.06$ mag) yield  a stellar mass of $M_*=3.9^{+0.9}_{-0.6}\times10^6 M_{\odot}$, which we also find to be consistent with near-infrared data. 
Via mass modelling, with our new measurements of velocity dispersion ($\sigma_{ap}=27\pm2$ km s$^{-1}$) and size ($R_e=20.9\pm1.0$ pc), 
we obtain an elevated dynamical-to-stellar mass ratio  $M_{dyn}/M_*=8.2$ (with a range $5.6\le M_{dyn}/M_* \le 11.2$).
Furthermore, we analyse the  surface brightness profile of S999, finding only a small excess of light in the outer parts with respect to the fitted S\'ersic profile, and a positive colour gradient.
Taken together these observations suggest that S999 is the remnant of a much larger galaxy that has been tidally stripped.  If so, the observed elevated mass ratio may be caused by mechanisms related to the  stripping process: the existence of an  massive central black hole or internal kinematics that are out of equilibrium due to the stripping event. Given the observed dynamical-to-stellar mass ratio we suggest that S999 is an ideal candidate to search for the presence of an overly massive central black hole.
\end{abstract}


\section{Introduction}

Ultra-compact dwarfs (UCDs) are a class of compact stellar system that was only discovered some  
15 years ago \citep{1999A&AS..134...75H,2000PASA...17..227D}. Their sizes (10 $\la R{_e}  \la$ 100 pc) and masses ($10^6 \la M/M_{\odot} \la10^8$) are intermediate between those of globular clusters (GCs) and compact ellipticals (cEs). Initially, they were 
called `galaxies' but today it is still debated whether UCDs are 
the high mass end of the globular cluster 
population \citep[e.g.][]{2002A&A...383..823M,2008MNRAS.389.1924F},
or are the surviving nuclei of galaxies that were torn 
apart by tidal forces \citep[e.g.][]{1994ApJ...431..634B,2001ApJ...552L.105B,2013MNRAS.433.1997P}. There is now growing evidence that both of these formation channels contribute to the UCD population
\citep[e.g.][]{Mieske:2006ia,2011AJ....142..199B,2011MNRAS.412.1627C,2011MNRAS.414..739N,2012MNRAS.422..885P}.

Recently \citet{2014MNRAS.443.1151N} presented sizes, stellar masses and internal velocity dispersion 
measurements for a large sample of UCDs. Velocity dispersions, coupled with size, provide an estimate 
of the dynamical mass that is relatively insensitive to the distribution of stellar orbits \citep{2010MNRAS.406.1220W}. Early studies, based on smaller samples of UCDs, detected a trend for the dynamical mass-to-light ratio (or equivalently the dynamical-to-stellar mass ratio) to increase steadily with increasing stellar mass 
\citep{2008MNRAS.389.1924F,2008MNRAS.386..864D,2008A&A...487..921M}. Using the largest sample to date of data from the literature, \citet{2014MNRAS.444.2993F} showed that this trend persisted for the general population of UCDs even up to stellar masses of $\sim$ 10$^9$ M$_{\odot}$. 

Several (non-exclusive) explanations for these elevated mass ratios have been offered,
including a non-canonical stellar initial mass function (IMF;  both a bottom or top-heavy IMF would lead to an underestimation of the stellar mass; e.g. \citealt{2008AN....329..964M,2009MNRAS.394.1529D,2010MNRAS.403.1054D,2012ApJ...747...72D}), the presence of dark matter
\citep[e.g.][]{2008MNRAS.391..942B,2008AN....329..964M}
and a central massive black hole raising the velocity dispersion \citep{2013A&A...558A..14M}. The latter 
possibility has been dramatically confirmed recently by \citet{2014Natur.513..398S} who detected an oversized central black hole in M60-UCD1 \citep{2013ApJ...775L...6S} that represents 15 per cent of the 
mass of the object. 
Although, interestingly despite the presence of this massive black hole M60-UCD-1 
 does not display a  strongly elevated mass ratio based on its unresolved velocity dispersion
  (e.g.\ \citealt{2014MNRAS.444.2993F} found a dynamical-to-stellar mass ratio of about unity).

\citet{2014MNRAS.444.2993F} also noted half a dozen UCDs with extremely high mass ratios (i.e.\ up to 10).
Such UCDs are of particular interest. They may represent a rare subpopulation of UCDs that 
have been caught in the early stages of the stripping process, having effectively inflated sizes (or even outer halo structures)  relative to their stellar mass. This may be the case for VUCD7 which 
reveals a core and halo structure in its surface brightness profile \citep{2008AJ....136..461E}. Using the size and 
stellar mass of the core gives a dynamical-to-stellar mass ratio that is $\sim3\times$ lower (although still higher than the typical UCD).
We also note the recent discovery of a halo around the Milky Way globular cluster NGC 1851 
\citep{2014MNRAS.442.3044M}. The presence of this halo of extra tidal light suggests that NGC 1851 was once the nucleus a dwarf galaxy, which was tidally stripped as it interacted  with the Milky Way. 

However, another more mundane possibility is that some of the measurements in the literature for these extreme objects are spurious. This appears to be the case for M59cO which had a high mass ratio based on the measurement of \citet{2008MNRAS.385L..83C} using a low resolution spectrum from the Sloan Digital Sky Survey. Obtaining a medium resolution spectrum from the Keck telescope yield a new velocity dispersion measurement, which \citet{2014MNRAS.444.2993F} showed gives its re-calculated mass ratio to be within the scatter of the UCD population as a whole.

\citet{2005ApJ...627..203H} studied a small number of UCDs around M87. They 
calculated a dynamical mass-to-light ratio for half a dozen objects, finding one (called S999) 
to have the highest ratio of $9.36\pm0.94$.\footnote{With the updated uncertainty of the velocity dispersion (see Section \ref{section:literature}) the uncertainty of the dynamical mass becomes $\pm1.97$.}
In the list of over 50 UCDs by \citet{2008A&A...487..921M}, the UCD with the highest mass-to-light ratio  
was also S999 with M/L$_V$ = 10.2. 
In the \citet{2014MNRAS.444.2993F} study, it has a similarly extreme dynamical-to-stellar mass ratio.
In both the  \citeauthor{2008A&A...487..921M} and \citeauthor{2014MNRAS.444.2993F} works the velocity dispersion, size and luminosity of S999 
are based on the measurements of  \citet{2005ApJ...627..203H}.

As well as an extreme mass ratio,
S999 also has a particularly low stellar mass and low density for a UCD. 
These extreme properties call for a confirmation of the mass ratio.
Here we take advantage of a very deep ACS pointing from the {\it HST} in two filters,  to remeasure the 
size and photometry of S999. In particular, we search for an extended halo in the surface brightness profile which may indicate ongoing stripping. We also present new 
high-quality Keck spectroscopy from which we obtain the velocity dispersion and stellar population properties (which in turn provide an improved estimate for the stellar mass).
We also analyse the surface brightness profile of seven other UCDs around M87 (within the 
deep ACS field of view) to search for extended halo light. 

\section{Observations and data reduction}

This study is based on very deep archival {\it HST}   images of S999 and seven more UCDs in the same field of view, and a newly obtained spectrum of S999 from the Keck telescope. In the following paragraphs the data acquisition and reduction are described. Here we assume that the UCDs are located at the same distance as M87,  for which surface brightness fluctuations and the tip of the red giant branch give consistent values, i.e.~16.7 Mpc  (\citealt{Blakeslee:2009dc,2010A&A...524A..71B}; see also \citealt{Anonymous:J4skqax8}),\footnote{In the literature often a distance of 16.5 Mpc was assumed for the M87 UCDs. Both the inferred physical size and stellar mass would be reduced by 1-3\%. In the mass ratio this  cancels partly, and the derived ratio for a distance of 16.5 Mpc would be larger by 1.2\%.}  which means 
that one arcsecond corresponds to $1\arcsec=80.35$ pc.

\begin{table}
\caption{M87 UCDs in the field of view of the ACS observations}
\begin{center}
\begin{tabular}{lccc}
\hline
Name & Right Ascension & Declination & $V_\textrm{hel}$ \\
 & $(h$:$m$:$s)$ & $(d$:$m$:$s)$ & $(km\ s^{-1})$ \\
\hline
    S999 &  12:30:45.91 &	+12:25:01.5 & $1466\pm5$\\
 H44905 &  12:30:57.08 &	+12:23:39.8 & $1563 \pm18$\\
   S672 & 12:30:54.73 & +12:21:38.3 &$735\pm106$ \\
    S887 &12:30:48.93 &	+12:21:55.6& $1811\pm 106$\\
    S928 &12:30:47.70&	+12 24 30.4&   $1283\pm5$\\
   S5065 &12:30:50.05&	+12:24:08.9&  $1578\pm3$\\
   S8005 &  12:30:46.20	&+12:24:23.1&$1883\pm5$\\
   S8006 &  12:30:46.65	&+12:24:22.2&$1079\pm5$\\
\hline\end{tabular}
\end{center}
The equinox of the coordinates is J2000. The data is taken from the compilation in \citet{2011ApJS..197...33S}, following their naming scheme with the \citet{2001ApJ...559..812H} extension of the \citet{1981ApJ...245..416S}  nomenclature.
\label{table:UCDs}
\end{table}

\subsection{{\it HST} ACS Imaging}
\label{sec:HST}

Very deep  archival  {\it HST} Advanced Camera for Surveys (ACS)
images are available in two filters (F606W and F814W, wide $V$ and $I$, respectively) from the {\it HST} GO programme 10543 (PI Baltz). 
While the original purpose of the observations was to study microlensing towards M87, the images have also been used to study its globular cluster system
\citep{Waters:2009kr,Peng:2009ew,Madrid:2009jx,Webb:2012cva}. 
Similarly, we used the depth and resolution of the observations to obtain surface brightness profiles of S999 and seven other UCDs  within this field of view. 
Furthermore, we processed the overlapping ACS {Virgo Cluster Survey} pointings (ACS VCS, \citealt{2004ApJS..153..223C}) F475W ($g$) and F850LP  ($z$) in the same fashion in order to enable a comparison of our measurements to  the ACS VCS study of compact stellar systems \citep{2005ApJ...627..203H}.

The pipeline processed {\it FLT.fits}  images were downloaded from the Mikulski Archive for Space Telescopes (MAST) archive\footnote{\url{http://archive.stci.edu}} and further reduced with the latest {\it HST} ACS pipeline {\it astrodrizzle}, similar to \citet{Madrid:2009jx}. The observations were carried out with a dither pattern allowing for subpixel drizzling, so that
the original pixel size of $0\farcs05$ was decreased to $0\farcs035$.
The drizzled and shifted images were combined with the IRAF task {\texttt imcombine}, the magnitudes were
calibrated to the Vega system using the provided zeropoints\footnote{\url{http://www.stsci.edu/hst/acs/analysis/zeropoints/zpt.py}} and corrected for Galactic extinction according to \citet{1998ApJ...500..525S}.\footnote{The difference in the attenuation as compared to \citet{2011ApJ...737..103S} is smaller than 0.015 mag in all bands.}

\begin{figure}
\includegraphics[width=0.48\textwidth]{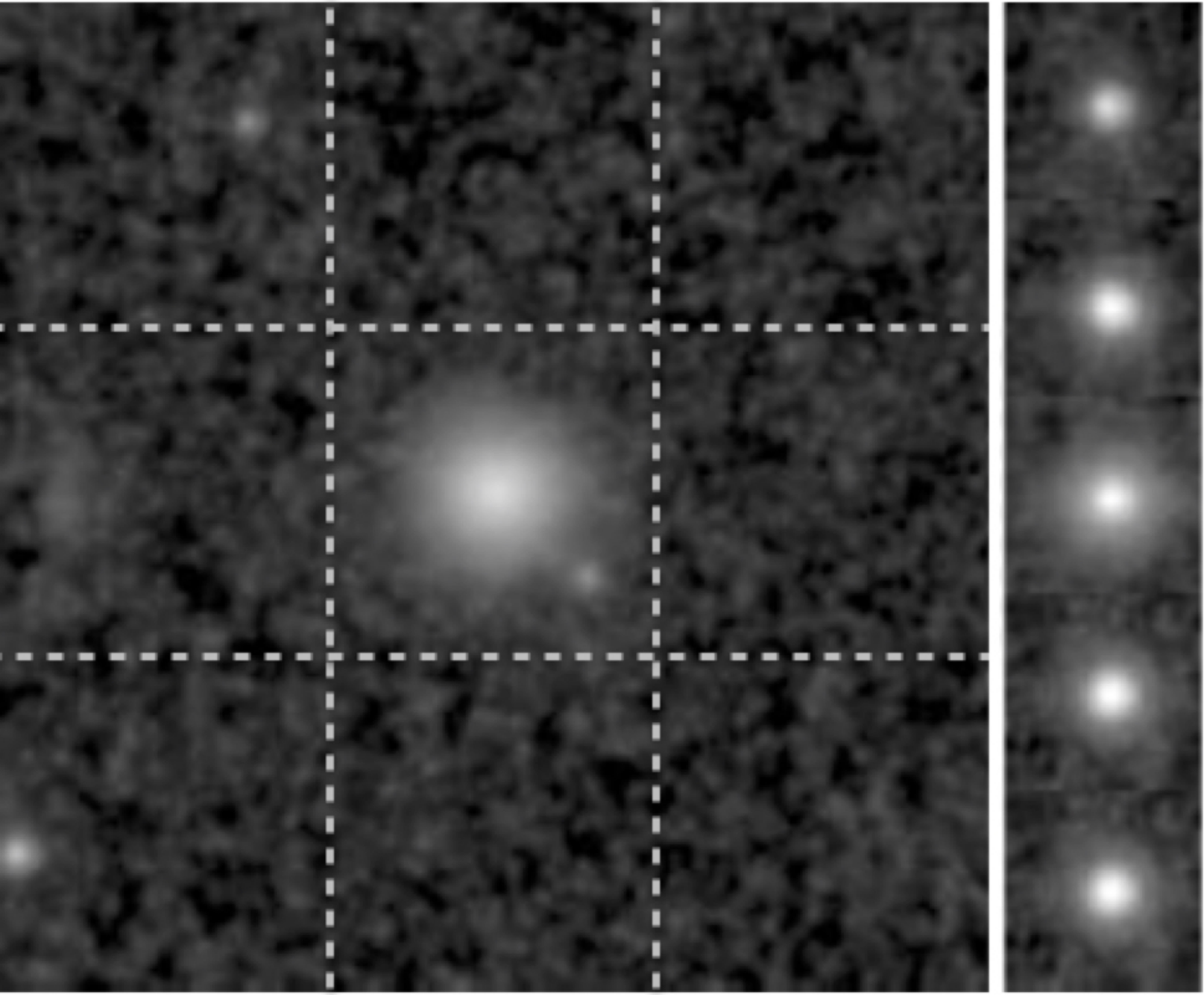}
\caption{Image of S999. {\it Left} panel: $F814W$-band image of S999 with the light of M87 subtracted. The panel displays $201\times201$ pixels or $\sim7\arcsec\times7\arcsec$; at the assumed distance of 16.7 Mpc, $1\arcsec = 80.35$ pc.  The grid indicates the eight surrounding boxes around S999 which were used to determine the local background values. {\it Right} panels: five globular clusters from \citet{2011ApJS..197...33S}, which are close to S999 in projection, shown on the same pixel and grey scale for comparison.}
\label{fig:S999+GC}
\end{figure}

The field of view is filled with the light from M87. 
In order to study the light distributions  of the UCDs, especially at faint surface brightness levels in their outskirts, the  light of M87 needed to be removed with high precision.
This was achieved by first subtracting from each pixel the median value of those in a $101\times101$ pixel box around that pixel \citep[see also][]{Madrid:2009jx}.
On this preliminarily processed image we ran {\sc Sextractor} \citep{1996A&AS..117..393B} to create an object mask. The final model of M87 was created by a combination of the IRAF tasks  \texttt{ellipse} and \texttt{bmodel} using that object mask. 
The final model 
was  subtracted from the image before any further analysis was carried out. 
We found a remaining non-zero local background around the UCDs. Therefore, we subtracted
the average of the mean pixel values in the surrounding eight boxes of a $3\times3$ grid with a total
width of 201 pixels centred on each UCD (see Fig.~\ref{fig:S999+GC}). The uncertainty of the background was estimated
by the scatter of these values and the mean values of 16 further boxes, which were placed 
at distances to the centre of M87 similar to those of the UCDs. 
An image of S999 and a few globular clusters in the surroundings are shown in Fig.~\ref{fig:S999+GC}.

\subsubsection{Surface brightness profiles}

\begin{figure*}
\includegraphics[height=0.9\textwidth,angle=-90]{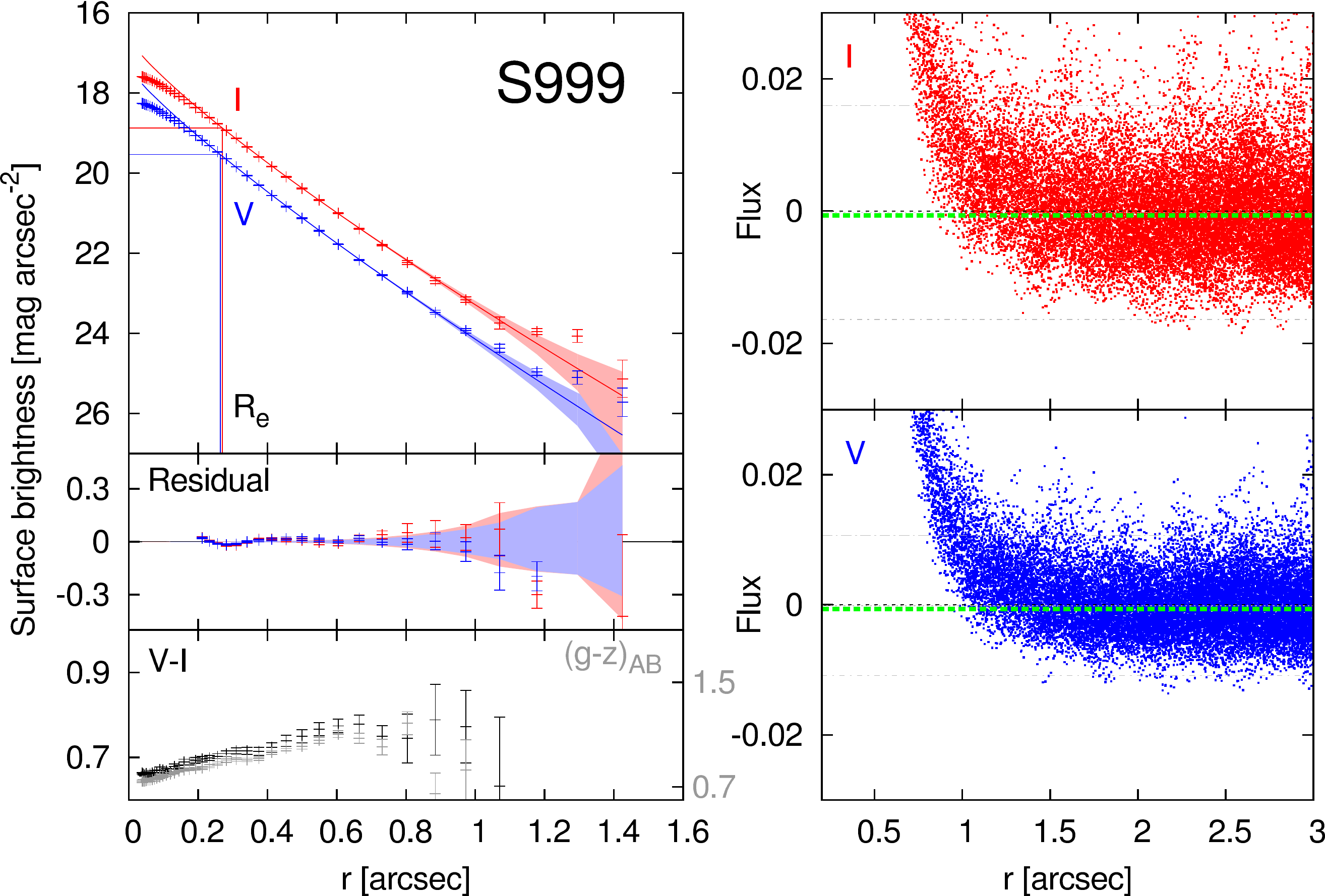}\\~\\

\caption{Surface brightness and colour profiles  of S999. {\it Left:} from {\it top} to {\it bottom} are  the surface brightness profiles and S\'ersic fits,  the residuals of the profile fits, and  the F606W-F814W ($V-I$, {black}) and F475W-F850LP ($g-z$) colour profiles (grey, with scale on right side).  Red and  blue colours represent F814W and F606W, respectively. The shaded areas in the upper two panels display the systematic uncertainty of the background determination, while the error bars display the measurement uncertainty as given by the {\tt ellipse} task of IRAF. 
The vertical and horizontal lines in the surface brightness profile panel indicate the half-light radius and the surface brightness at this radius  (at the assumed distance of 16.7 Mpc, $1\arcsec = 80.35$ pc).
{\it Right:} Pixel values on a linear scale for F814W \textit{(top)} and F606W \textit{(bottom)} to assess the background subtraction, with the {green} horizontal line displaying the average pixel value at radii between $2\arcsec$ and $3\arcsec$ after a $2.3 \sigma$ clipping. \label{fig:profile}}
\end{figure*}

The surface brightness profiles of S999 and seven more UCDs in the ACS field of view (Fig.~\ref{fig:profile} and \ref{fig:profiles}) were measured with the \texttt{ellipse} task of IRAF \citep{1987MNRAS.226..747J}.
During the process fore- and background objects were masked. 
The position angle and ellipticity of the isophotes were fixed to values obtained by averaging the isophotes with semi-major axis lengths between $0\farcs25$  and $0\farcs6$ from a preliminary profile measurement (with the two quantities as free fitting parameters). The spatial range was chosen such that the measurements were unaffected by the point spread function (PSF) and benefited from a high signal-to-noise.
The UCDs are all quite round (axis ratios $b/a$ from 0.84 to 0.97).
However, the isophotes in the outskirts can be severely biased by extreme pixel values due to noise in the images.
The not well-constraint isophotes vary unreasonably to random values of position angle and ellipticities, and especially the latter biases the surface brightness profile.
For our specific question -- looking for signs of extended envelopes in the outer parts of the UCDs -- this was much less acceptable
than the potential loss of information from fixing position angle and ellipticity.
The azimuthally averaged  surface brightness profiles reach typically a surface brightness in the $I$-band of $\sim 24$ mag arcsec$^{-2}$ at 3 to 5 half-light radii. 

We fitted S\'ersic profiles to the measured surface brightness profiles along the semi-major axes using a nonlinear least-squares Levenberg-Marquardt algorithm,
weighting the individual measurements with the inverse squared of the uncertainties as provided by the ellipse task.
For the fits we excluded the innermost $0\farcs15$ so as not to be affected by the PSF
(the typical full-width at half maximum of the PSF for the ACS wide-field camera is $\sim0\farcs1$). 

\subsubsection{Photometry}
The photometry for the eight UCDs in the ACS field of view was obtained in two different ways:
a non-parametric approach and via  fitted S\'ersic functions.

For the non-parametric approach we followed \citet{2008ApJ...689L..25J} and determined
(elliptical) Petrosian apertures \citep{1976ApJ...209L...1P}. 
The total flux and half-light semi-major axis length $a_e$ were then measured in an aperture with twice the Petrosian 
semi-major axis length, and corrected for the missing flux following the prescriptions of \citet{2005AJ....130.1535G}.
The parametric photometry was calculated from the S\'ersic fits to the light profiles using the formulae
provided in \citet{2005PASA...22..118G}.

Our photometric measurements are summarised in Table~\ref{table:photometry}.
The deviations between non-parametric and parametric photometry can be attributed to effects of the PSF, as
the two objects  with the greatest difference are those with the largest S\'ersic $n$, and hence the steepest profile in the innermost region. 
The uncertainty on the individual measurements is small, which can be seen by the consistency 
 of the systematic increase of size with wavelength along the sequence $g$-$V$-$I$-$z$ with the positive colour gradient of S999 (Tables~\ref{table:photometry} and \ref{table:ACSphotometry}).
Furthermore, the $V-I$  colours inferred from the total magnitudes from the two methods of measuring photometry are consistent.
Our measurements on the ACS VCS images are very consistent with those of \citet{2005ApJ...627..203H} using the same data: the typical deviation for the magnitudes and radii derived from the profile fits (\citeauthor{2005ApJ...627..203H} used King profiles) for the three UCDs in common is at the 5\% level. In the subsequent analysis we consider this as the uncertainty level for the luminosities and radii (see also Table~\ref{table:S999lit}), and adopt the average circularised radius from the two S\'ersic fits as the size $R_e=20.9$ pc.

\begin{table*}
\caption{Photometry of UCDs around M87}
\begin{center}
\begin{tabular}{l|cc|cc|cccc|cccc|cc}
\hline
UCD & \multicolumn{4}{c}{Non-parametric}  & \multicolumn{8}{c}{S\'ersic fit} &$b/a$ & {PA} \\
 & \multicolumn{2}{c}{F606W ($V$)}  & \multicolumn{2}{c}{F814W ($I$)}  & \multicolumn{4}{c}{F606W ($V$)}  & \multicolumn{4}{c}{F814W ($I$)}\\
 & $m$ & $a_e$ [$\arcsec$] & $m$ & $a_e$ [$\arcsec$]  & $m$ & $a_e$ [$\arcsec$] & $\mu_e$ & $n$ & $m$ & $a_e$  [$\arcsec$]& $\mu_e$ & $n$ & & [$^\circ$] \\ 
\hline
    S999 &  19.78 &  0.29 & 19.05 &  0.30  &  19.71 &  0.26 & 19.54&    1.2  & 18.98 &  0.27 & 18.88 &   1.2 & 0.96 & 8.8 \\
  H44905 &  21.38 &  0.27 & 20.65 &  0.28  &  21.32 &  0.25 & 20.94&    0.9  & 20.62 &  0.26 & 20.28 &   0.8 & 0.97 & -53.1 \\
    S672 &  20.34 &  0.24 & 19.62 &  0.24  &  20.04 &  0.17 & 19.20&    2.1  & 19.43 &  0.20 & 18.78 &   1.8 & 0.94 & -30.0 \\
    S887 &  20.60 &  0.18 & 19.82 &  0.19  &  20.12 &  0.11 & 18.49&    2.6  & 19.45 &  0.13 & 18.09 &   2.4  & 0.96 & -54.6\\
    S928 &  19.27 &  0.34 & 18.58 &  0.35  &  19.19 &  0.30 & 19.38&    1.5  & 18.52 &  0.31 & 18.75 &   1.4 & 0.92 & -6.7 \\
   S5065 &  19.67 &  0.21 & 18.93 &  0.22  &  19.45 &  0.17 & 18.43&    1.2  & 18.74 &  0.18 & 17.81 &   1.2 & 0.97 & 27.9 \\
   S8005 &  19.96 &  0.39 & 19.25 &  0.39  &  19.94 &  0.36 & 20.31&    1.2  & 19.26 &  0.36 & 19.62 &   1.2  & 0.84 & 16.6\\
   S8006 &  20.02 &  0.28 & 19.32 &  0.28  &  19.91 &  0.24 & 19.64&    1.3  & 19.19 &  0.24 & 18.90 &   1.3  & 0.96 & 16.4\\
     \hline\end{tabular}
\end{center}
The semi-major axis length $a_e$ of the apertures containing half of the light is given in arcseconds and the surface brightness $\mu_e$ of the corresponding isophote in mag arcsec$^{-2}$; $m$ is the total magnitude. The final two columns list the  minor-to-major axis ratio ($b/a$) and position angle (PA) in degrees.
At the assumed distance of 16.7 Mpc, one arcsecond corresponds to $1\arcsec = 80.35$ pc. 
The photometry based on the ACS VCS pointing is summarised in Table \ref{table:ACSphotometry}.
\label{table:photometry}
\end{table*}

\subsection{Near-infrared photometry of S999}
In addition we measured the $K_{\rm s}$-band magnitude of S999  from archival WIRCam \citep{2004SPIE.5492..978P}
imaging of M87 taken as part of the Next Generation Virgo Cluster Survey-IR \citep{Munoz:2014ea}. 
The final calibrated $K_{\rm s}$-band stacked image of M87 was downloaded from the CADC archive, this had a total
integration time of 4487s and provided image quality of 1.31 arcseconds; sufficient for the measurement of a
total integrated magnitude of S999 but not for any examination of spatially resolved properties. The downloaded
image was first analysed to remove the halo of M87 using the procedure described in \citet{2011MNRAS.414..739N}. 
The total integrated $K$-band magnitude of S999 was then measured with {\sc SExtractor} \citep{1996A&AS..117..393B} yielding a magnitude of 
17.71 $\pm$ 0.03 or $M_{K_s} = -13.40\pm0.03 $ mag.

\subsection{Keck ESI Spectrum of S999}
A high-quality spectrum for the UCD S999 was observed during the night of the 2014 March 21 with the 
{Echelle Spectrograph and Imager} \citep[ESI,][]{Sheinis:2002ft} with the Keck II telescope with a total integration time of 3.2 hours, resulting in a signal-to-noise of $S/N\sim20$ per \AA{}ngstr\"om.
The centre of the UCD was placed in the middle of the slit, which was aligned with the parallactic
angle (PA$=-73^\circ$). 
In total seven exposures were obtained (one with an integration time of 600s  and six with 1800s)
in the echellette mode using a $0\farcs75\times20\arcsec$ slit, with the seeing varying between $0\farcs5$ and $0\farcs6$.
The setup results in
a resolving power of $R\sim5400$. 
The ten Echelle orders span a wavelength range of approximately 3900 to 11000 \AA.
The spatial scale corresponds to $0\farcs12$ to $0\farcs17$ per pixel from the blue to red end of the spectrum.

From the several  frames for each of the calibrations (internal flat fields, bias, arc lamps) 
we determined master calibration frames by combining the two-dimensional spectra using 
the {\tt imcombine} task in IRAF.
The individual science spectra were reduced with the pipeline {\sc makee} (T.~Barlow).
The software takes care of the flat fielding and bias subtraction. It also uses a spectrum of a star to 
trace the Echelle orders in order to extract a one-dimensional spectrum for each order, which is then
wavelength calibrated using the known lines of the arc lamps (of a previous observing run using a similar setup). 
 Also an error spectrum, i.e.\ the individual uncertainties of the pixels in the spectrum, is calculated.

The flat fielding does not bring the various Echelle orders to a common flux level. Therefore, we took
the spectrum of a velocity standard (HR3454) observed with the same setup and obtained the blaze  function by comparing that spectrum
to a published spectrum of the same star \citep{2003A&A...402..433L}.
The Echelle orders of the corrected spectrum overlap. 
The various exposures and different Echelle orders are finally combined to a single spectrum in a signal-to-noise optimised
way using UVES\_popler (M.T.~Murphy).  The uncertainties are propagated through the scaling and combination processes in order to obtain a final error spectrum.

The ESI spectrum of S999 was fitted with a penalised pixel fitting code, 
{\sc pPXF} \citep{2004PASP..116..138C}, using  stellar spectra of the \textsc{elodie} library 
\citep[][version 3.1]{2001A&A...369.1048P}. The library contains spectra of 1388 stars, which cover  a wavelength region of 3900 \AA{} to 6800 \AA{} at a resolution of $\sigma\sim$12 km s$^{-1}$  (R$\sim$10,000), and which span a wide range of spectral type and metallicity. For the fitting, the template stars were convolved to the instrumental resolution (with a constant $\sigma=21$ km s$^{-1}$), taking into account the templates' resolution. The spectral range used for the fitting was $4050-5550$ \AA{}.
The observed and best-fit model spectra are shown in Fig.~\ref{fig:bestfit}.

To estimate the uncertainties for both of the velocity dispersion and the Lick indices (see below), we performed Monte Carlo simulations: 
The observed spectrum was altered with random modifications drawn from a normal distribution with a width according to the error spectrum. The modified spectrum was then fitted in the same way, and the steps were repeated 1000 (for the velocity dispersion) or 500 (for the Lick indices) times. The adopted values and their uncertainties are then given by the mean and standard deviation of Gaussian fits to the distribution of the individual fitted values.

\subsubsection{Internal kinematics}
\begin{figure}
\includegraphics[width=0.5\textwidth]{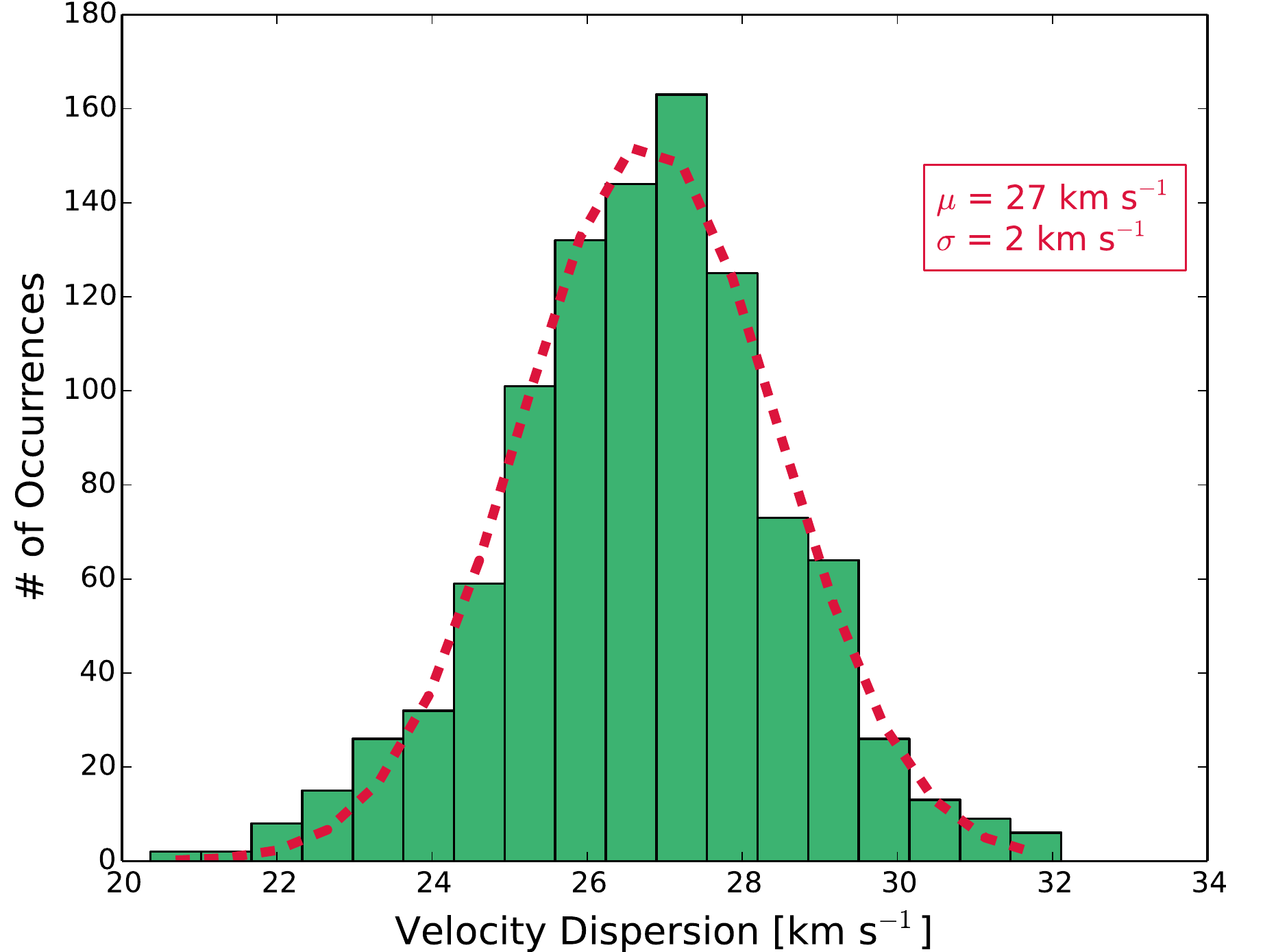}
\caption{
Distribution of velocity dispersion values of S999 from  1000 Monte Carlo simulations, simulating random artificial noise according to the error spectrum.  We adopt the  values for the  velocity dispersion and its uncertainty as the centre and sigma (see box) from a Gaussian fit to the distribution (red dashed line).}
\label{fig:mc}
\end{figure}

The recession velocity and velocity dispersion were determined with the best-fitting template by \textsc{pPXF}. 
 We discarded the calcium (Ca-)triplet lines due to a lack of a sufficiently high resolution stellar library for this spectral range. For the final velocity dispersion and its uncertainty 
we adopted from the Monte Carlo simulation of the fitting process, i.e.\ $\sigma_{ap}= 27\pm2$ km s$^{-1}$.
Therefore, we obtained a realistic estimate for the uncertainty in the velocity dispersion  as demonstrated in Fig.~\ref{fig:mc}.
Also, systematic errors from template mismatch were minimised by using a large library of template stars.
Using  velocity standard stars (a small number from our observing run, and those from \citealt{2005ApJ...627..203H}) observed with the same setup,
for a similar wavelength range as well as for the Ca-triplet, led to consistent velocity dispersions but larger uncertainties (for comparisons with the \citeauthor{2005ApJ...627..203H} data see Section \ref{section:literature}).
We also note that the measured velocity dispersion is well above the instrumental resolution and
\citet{geha} found that it is feasible to reliably measure (with an accuracy of 10\%) velocity dispersions  ($\sigma\sim18.5$ km s$^{-1}$) below the instrumental resolution with a similar setup and lower signal-to-noise ratio.

\defcitealias{2005ApJ...627..203H}{H05} 	  	 
\defcitealias{2011ApJS..197...33S}{S11}		
\defcitealias{2013A&A...558A..14M}{M13}

\begin{table}
\caption{Literature values and our measurements for S999}
\begin{center}
\begin{tabular}{lcc}
\hline
Quantity & Literature & This study \\
\hline
\multicolumn{3}{l}{Magnitude}\smallskip \\
$M_{g,AB}$ & $-10.791$  \citepalias{2005ApJ...627..203H} &  $ -10.79 \pm  0.06 $ \\    
$M_{i,AB}$ & $-11.44$ \citepalias{2011ApJS..197...33S} & --- \\   			    
$M_{z,AB}$ & $-11.711$  \citepalias{2005ApJ...627..203H} & $-11.69  \pm 0.06 $\\      
$M_V$ & $-11.16$ (M13) &  --- \\  
$M_{F606W(V)}$  & --- & $-11.40 \pm 0.06$\\ 
$M_{F814W(I)}$  & --- & $-12.13 \pm 0.06$\\
$M_{K_s}$ & --- & $-13.4\pm 0.03$\\
\hline 
\multicolumn{3}{l}{Size [pc]} \smallskip \\
$R_{e,g} $  & $19.9 \pm 0.2$ \citepalias{2005ApJ...627..203H} & $20.2 \pm 1.0$\\
$R_{e,i} $  &   34.1 \citepalias{2011ApJS..197...33S} & --- \\  	
$R_{e,z} $  & $21.9\pm 0.5$ \citepalias{2005ApJ...627..203H} &  $22.0 \pm 1.1$\\ 
$R_{e,F606W(V)}$ & --- &  $20.8\pm1.0$ \\ 
$R_{e,F814W(I)}$  & --- &  $21.3\pm1.1$ \\
\hline
\multicolumn{2}{l}{Velocity dispersion  [km s$^{-1}$]}\smallskip \\
$\sigma_{ap}$& $23.3 \pm 1.3$ \citepalias{2005ApJ...627..203H} & $27 \pm 2$\smallskip\\
$\sigma_{tot} $ & --- & $26\pm2$ \\
$\sigma_0 $ & $ 25.6 \pm 1.4$ \citepalias{2005ApJ...627..203H}& --- \\
$\sigma_0 $ & $23.2 \pm 1.3$ (M13) &---\\
$\sigma_0 $ & $ 26.2 \pm 1.3$ (MB14)&--- \\
\hline
\multicolumn{3}{l}{Metallicity}\smallskip \\
$[\textrm{Fe}/\textrm{H}]_{gz}$ & $-1.93$ \citepalias{2005ApJ...627..203H} & ---\\
$[\textrm{Fe}/\textrm{H}]_{CT_1}$ & $-1.38$ \citepalias{2005ApJ...627..203H} & ---\\
$[\textrm{Fe}/\textrm{H}]$ & $-1.40$  (M13) & ---\\
$[\textrm{Fe}/\textrm{H}]_{spec}$  & ---& $-1.27\pm0.12$\\
\hline
Age [Gyr]  & --- & $7.6^{+2.0}_{-1.6}$\\
\hline
Mass  [10$^6$ $M_{\odot}$]  \smallskip\\
 $M_{dyn}$ & $24.3 \pm 2.7$ (M13) &   \multirow{3}{*}{ $32\pm 5$} \\ 
 $M_{dyn}$ & $22.4\pm 3.0$ \citepalias{2005ApJ...627..203H}\footnotemark &\\ 
 $M_{dyn}$  & 21.2 (F14)&\smallskip \\ 		
$M_{*}$ & 2.6 (F14) &   \multirow{2}{*}{ $3.9^{+0.9}_{-0.6}$}\\  
$M_{*}$ & 5.6 (M13) & \\					
\hline
 \multicolumn{3}{l}{Ratios}\smallskip \\
$M_{*}/L_z$  & ---& $1.3^{+0.3}_{-0.2}$ \smallskip \\  
$M_{dyn}/M_{*}$ & $8.5$ (F14) &   \multirow{2}{*}{5.6...{\bf 8.2}...11.2} \\   
$M_{dyn}/M_{*}$ & $4.45 \pm 0.87$ (M13) & \\   
  \hline
  \end{tabular}
\end{center}
The references in the Table are: H05 -- \citealt{2005ApJ...627..203H}; S11 -- \citealt{2011ApJS..197...33S}; M13 -- \citealt[using H05 data]{2013A&A...558A..14M}; MB14 -- Mieske \& Baumgardt 2014, priv.\ comm.; F14 -- Forbes et al. 2014; the values were converted to our assumed distance of 16.7 Mpc when necessary. 
The velocity dispersion $\sigma_{tot}$ is from the mass modelling, integrating over the whole object.
For the mass ratio we list our preferred value (bold), as well as the range from lower/upper  estimates for dynamical and stellar mass.
\label{table:S999lit}
\end{table}
\footnotetext{Obtained via a fit of a King model.}

\subsubsection{Stellar populations}
In order to derive stellar population parameters,  we measured Lick indices according to the definition of \citet{1998ApJS..116....1T}. The measured indices are corrected to the Lick resolution
by multiplying the index measured on the observed spectrum $I_{\mathrm{Obs}}$
with the ratio of the index measured 
on the best-fit model convolved to the observed velocity dispersion  ($I_{\mathrm{best}}^{\mathrm{Obs}}$)  and the Lick resolution  ($I_{\mathrm{best}}^{\mathrm{Lick}}$):
\begin{equation}
I_{\mathrm{Corr}}\,=\,I_{\mathrm{Obs}}\,\frac{I_{\mathrm{best}}^{\mathrm{Lick}}}{I_{\mathrm{best}}^{\mathrm{Obs}}}.
\end{equation} 
The Lick resolution for each index is taken from \citet{2005ApJS..160..163S}, and for the convolution the template resolution with a FWHM of 0.55  \AA{} was subtracted in quadrature.

We interpolated the  \citet{2011MNRAS.412.2183T} single stellar population (SSP) models to a fine grid (steps of 0.02 dex in each direction; \citealt{Onodera:2014un}) of age, metallicity ([Z/H]), and $\alpha$-abundance ([$\alpha$/Fe]) in order to find the SSP-equivalent stellar population parameters of S999 via $\chi^2$ minimisation,
\begin{equation}
\chi^2\,=\,\frac{(I_{\mathrm{Thomas}}\,-\,I_{\mathrm{Measured}})^2}{\sigma^2_{\mathrm{Index}}}
\end{equation}
where $\sigma_{\mathrm{Index}}$ is the $1\sigma$ uncertainty of the index derived from Monte Carlo simulations.

\begin{figure*}
\includegraphics[width=\textwidth]{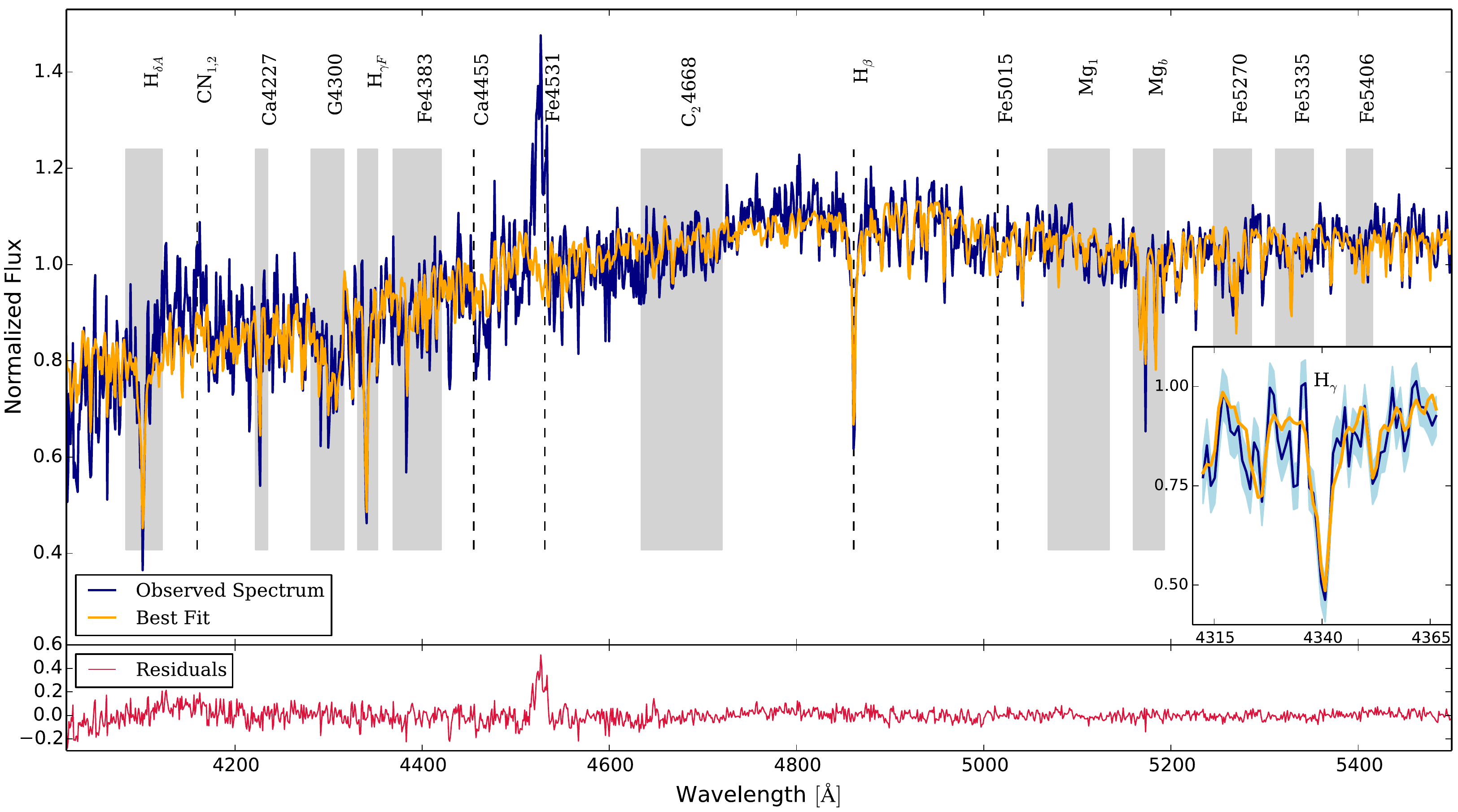}
\caption{Spectrum of S999 in the rest frame. The {\it top} panel shows the observed spectrum  (blue line, shifted to the rest frame) and the bestfit template obtained with \textsc{pPXF} (orange line).  
The grey shaded regions indicate the pass bands of indices used in the determination of age and metallicity (omitting $H_{\gamma_A}$ and {\it Mg}$_2$, which overlap with $H_{\gamma_F}$ and {\it Mg} $b$, respectively), while the remaining indices are indicated with the dashed horizontal lines.
A zoom in on the H$_\gamma$ line in the inset shows the quality of the fit, with the light blue shaded area indicating the uncertainty of the observed spectrum. The residuals are shown in the {\it bottom} panel (red line). The residual  around 4530 \AA{} is caused by a bad column in the detector. We exclude the Fe4531 index, when fitting for the stellar population parameters, since it coincides with some faulty pixels of the detector.}
\label{fig:bestfit}
\end{figure*}

The indices included in the analyses were  H$\delta_\mathrm{A}$, Ca4227, G4300, H$\gamma_\mathrm{A}$,
 H$\gamma_\mathrm{F}$, Fe4383, C$_2$4668, Mg$_1$, Mg$_2$, Mg$b$, Fe5270, Fe5335, and Fe5406.  The other indices in the wavelength range
 (H$_{\delta_{F}}$, CN1, CN2, Ca4455, Fe4531, H$_{\beta}$,   Fe5015, Fe5709, Fe5782, NaD, TiO$_1$, TiO$_2$) were excluded since they are either not well calibrated in the model or significant outliers in the model comparisons.
 In the comparison of observed and bestfit spectrum (Fig.~\ref{fig:bestfit}) there are some notable broader deviations (e.g.~around 4150 \AA{}, 4600 \AA{}, and 4800 \AA{}). These are caused by combining the various Echelle orders with imperfect corrections for the blaze function, resulting in a slight mismatch of the spectra's slopes in the overlap region of a pair of subsequent orders. 
 It is possible that only  the C$_2$4668 and H$_{\delta}$ indices are affected. Omitting those or using a slightly different blaze correction obtained from a different star do not change any of our conclusions. 
 
 We obtain: Age $=7.6^{+2.0}_{-1.6}$ Gyr; $[Z/\textrm{H}]=-0.95^{+0.12}_{-0.10}$;  $[\alpha/\textrm{Fe}]=0.34^{+0.10}_{-0.12}$;  $[\textrm{Fe}/\textrm{H}]=-1.27\pm0.12$ (assuming  $[\textrm{Fe}/\textrm{H}]=[Z/\textrm{H}]-0.94 [\alpha/\textrm{Fe}]$, \citealt{Thomas:2003kx}). The predicted model colours $F606W(V)-F814W(I)=0.72$  and $(g-z)_{AB}=1.04$ for an age of 8 Gyr and metallicity $[Z/\textrm{H}]=-0.95$  are in good agreement with the  observed integrated colours of  $F606W(V)-F814W(I)=0.73\pm0.08$ and $(g-z)_{AB}= 0.90\pm0.08$ (Fig.~\ref{fig:profile}).

The colour gradient is a hint that the stellar population of S999 may not be well represented by the assumption of a simple stellar population.
In order to estimate the effect on the stellar mass-to-light ratio of a more complex star formation history we used a model grid with the combination of two simple stellar populations:
we combined the SSPs of  \citet{2011MNRAS.412.2183T} with the model of a population  of an age of 15 Gyr, $[Z/\textrm{H}]=-2.25$, and $[\alpha/\textrm{H}]=0.3$, with a mass fraction of 10\% to 90\% in steps of 10\%.
The indices in the combined model were calculated as luminosity weighted averages using the mass-to-light ratios of \citet{2005MNRAS.362..799M}.
The smallest $\chi^2$ was achieved with half of the mass in each of the populations. The young population was then fitted with an age of $4.5^{+3.5}_{-1.7}$ Gyr, $[Z/\textrm{H}]=-0.10^{+0.30}_{-0.25}$, and $[\alpha/\textrm{Fe}]=0.10^{+0.10}_{-0.15}$. The total stellar mass for this model is slightly larger than the SSP-equivalent, but not by a lot: $M_*=4.1\times10^6 M_{\odot}$ and actually within the SSP stellar mass uncertainty.

\section{Results and Discussion}

\subsection{Light profile}
All eight analysed UCDs follow single S\'ersic profiles  quite well, although two of them show deviations   at a surface brightness around 20 to 21 mag arcsec$^{-2}$ in the $I$ band (see also Appendix~\ref{sec:profiles}).
S999 is the only object with a small excess of light in the outer parts, which is significant in both bands
after taking measurement and background subtraction uncertainties into account (Fig.~\ref{fig:profile}).
However, the excess is at a low surface brightness level of  $\sim24$-25 mag arcsec$^{-2}$ in $I$ and $V$, respectively (the excess contributes only a small fraction of the total light, i.e.~$\sim2$\%).
For comparison, the two S\'ersic components  of VUCD7 \citep{2008AJ....136..461E} 
contribute equally to the surface brightness of $\mu_V\approx20.6$ mag arcsec$^{-2}$ at $r\approx0\farcs4$.

The size of S999 based on the S\'ersic fit to the surface brightness profile might be underestimated, since the image was not deconvolved, even though the innermost region was excluded from the fit. However, the obtained size  is consistent with literature values (see Table~\ref{table:S999lit}), with the exception of \citet{2011ApJS..197...33S}, who fitted a King model with fixed concentration. 

At least three UCDs -- S999, S887, and S5065 -- have colour profiles that become steadily bluer towards larger radii over a range with a reliable colour measurement, by about $\sim$ 0.15 mag. 
Again, the uncertainty of the background subtraction is the major limitation. However, the colour profiles are consistent with the $g-z$ profiles measured from the ACS VCS images. 
 \citet{2008AJ....136..461E} found similar positive colour gradients for the brightest UCDs in their sample.

We note that the fourth cosine term profile, i.e.~disky or boxy isophotes, does not reveal any significant features correlated with the surface brightness profile of S999 (not shown).

\subsection{Extreme mass ratio}
 We determined the dynamical mass of S999 by assuming it is perfectly described as a S\'ersic function. 
Since the size has a slight wavelength dependence, we used an average value of those radii from the S\'ersic fits to surface brightness profiles in $F606W$ and $F814W$ ($R_e=20.9$ pc).
Using the equations in \citet{ciotti1991}, we deprojected the observed S\'ersic function to create a smooth three-dimensional density distribution. We assumed an initial mass normalisation and calculated the resulting projected velocity dispersion as a function of projected radius. From the density distribution we sampled $10^6$ stars, calculated their projected radii, and assigned each star a velocity along the line of sight consistent with the projected dispersion at that projected radius. We convolved each star with a Gaussian assuming a FWHM of the seeing ($0\farcs6 \sim  48.6$ pc). We then projected the angular dimensions of the slit ($0\farcs75 \times 2\farcs5$) to the distance of S999 and integrated each stellar seeing disk over the slit.  We next calculated the observed  integrated velocity dispersion within the slit, using the slit integration factors as weights in the integrated dispersion. Finally, we scaled the mass normalisation appropriately until the modelled integrated velocity dispersion $\sigma$ within the slit matched the observed value.\footnote{Our mass modelling is basically following \citet{2007A&A...463..119H}. However, they truncated the profiles at 500 pc (H.~Baumgardt, priv.\ comm.). The resulting dynamical masses are slightly smaller than they would be without the truncation, with the difference depending on the profile. For example for the Fornax UCD4 ($R_e=24.1$ pc, $n=5.5$, \citealt{2007AJ....133..1722E}) it is a 3\% effect.}
The resulting total dynamical mass is $M_{dyn}=32\pm5\times10^6 M_{\odot}$.
Given the large number of stars used in the simulation, we found that the random uncertainty in the total dynamical mass is dominated by the uncertainty in the measured velocity dispersion and that sampling uncertainties are not important.

For the conversion from luminosity to stellar mass
we took the stellar mass-to-light ratios assuming a Kroupa IMF from \citet{2005MNRAS.362..799M},  which are valid for the  \citet{2011MNRAS.412.2183T} stellar population models, which we used.\footnote{Which horizontal branch morphology (blue or red) is assumed for the lowest metallicity models has no influence on our mass-to-light ratios in the  $z$- and $K_s$-bands.}
We chose to use the $z$-band,  since the mass-to-light ratio
depends less critically on the age, $[Z/\textrm{H}]$, and $[\alpha/\textrm{Fe}]$ of the stellar populations
 at longer wavelengths.
The uncertainty of the total luminosity in the $z$-band with lower $S/N$ as compared to e.g.~$F814W$ is still small, and the model $M/L$ is readily available.
 Interpolating the model stellar mass-to-light ratios resulted in approximately $M/L_z=$ 1.1 to 1.6 for our observed age and metallicity including the ranges of uncertainty. The best guess is $M/L_z = 1.3$, which results in a stellar mass of $M_*=3.9 \times 10^6 M_{\odot}$ ($M_* =$ 3.3 to 4.8$  \times 10^6 M_{\odot}$ for the range of mass-to-light ratios).

Furthermore, we converted the ground-based $K_{\rm s}$ magnitude  into stellar mass, benefitting from the even longer wavelength.
With the measured age and metallicity,  \citet{2005MNRAS.362..799M} yielded  $M/L_{K_{\rm s}} = 0.8\pm0.1$ (the range comes again from the uncertainties in age, $[Z/\textrm{H}]$, and $[\alpha/\textrm{Fe}]$) and a stellar mass of 
$M_*=3.8^{+0.6}_{-0.5}\times10^6 M_{\odot}$. Using the \citet{2012MNRAS.427..127B,2013EPJWC..4303001B} models and the procedure  described in \citet{2014arXiv1407.6005N}
gave a slightly larger, but consistent mass of $M_*=4.7^{+0.9}_{-0.6}\times10^6 M_{\odot}$.
In both cases the uncertainties are dominated by those of the measured stellar population age.
Both measurements  are in good agreement with the stellar mass derived from {\it HST} imaging ($3.9\times10^6 M_{\odot}$).

The resulting dynamical-to-stellar mass ratio  is  $\sim$8.
The range of possible mass ratios is rather large, due to the uncertainties in both the 
stellar mass-to-light ratio and the dynamical mass estimate (i.e.\ 5.6 to 11.2 with the lowest dynamical and highest stellar mass and vice versa).
However, the mass ratio remains elevated even when adopting a stellar mass at the high
end ($M_{*}=4.8\times10^6 M_{\odot}$) of the possible range and scaling the dynamical mass down to the lowest published velocity dispersion ($\sigma=20$ km s$^{-1}$; \citealt{2005ApJ...627..203H} using \texttt{fxcor}), i.e.\ $M_{dyn}/M_{*} \ge 3.7$. 

\subsection{Comparison to literature}
\label{section:literature}
An extreme mass ratio  has previously been calculated for S999 \citep{2005ApJ...627..203H,2008A&A...487..921M,2014MNRAS.444.2993F}, all based on the ESI spectrum of  \citeauthor{2005ApJ...627..203H}.
With improved photometry from deeper images (factor $\sim$60 in total integration time),
and new spectroscopy (factor $\sim$5 in total integration time) allowing for additional constraints
 on the stellar population we confirm that the mass ratio is indeed elevated. 

\citet{2005ApJ...627..203H} measured the velocity dispersion of S999 in two different ways: with cross-correlation and with \textsc{pPXF}. Their two measurements are not compatible with each other ($\sigma_{ap}=19.6\pm1.0$ km s$^{-1}$ and $\sigma_{ap}=25.6\pm1.4$ km s$^{-1}$, respectively), but this might be explained by an underestimation of the uncertainty. 
 We note that for both their methods \citeauthor{2005ApJ...627..203H} used five stars, which were observed with the same setup. Here we made use of the much larger library of high resolution ELODIE stellar templates, which were observed with a different instrument. 
Repeating their measurement with \textsc{pPXF} around the magnesium triplet region, we reproduce their velocity dispersion, but obtain a 2-5 times larger uncertainty $\sigma_{ap}=24.5\pm3.5$ km s$^{-1}$, which is not accounting for systematic uncertainties. 
 The  velocity dispersion measurement using our own spectrum agrees well with that of \citeauthor{2005ApJ...627..203H} when using  \textsc{pPXF} (as we do). 
 Given our uncertainty estimate, which is larger than that of \citeauthor{2005ApJ...627..203H} for their spectrum
with  5 times shorter total integration time, our updated larger estimate for the uncertainty of the data with a smaller signal-to-noise ratio seems consistent, showing the usefulness of the Monte Carlo simulation for a realistic estimate.

Our size measurements for the ACS VCS pointings are in good agreement with those of \citet{2005ApJ...627..203H}, even though our approach was  simpler in the sense that we excluded the innermost region from the fit of the surface brightness profile instead of taking into account the convolution by the PSF as \citeauthor{2005ApJ...627..203H} did.
Also, the radii measured on the deeper images are very consistent. All of the measurements follow the mild trend to larger radii at longer wavelengths, consistent with the colour gradient of S999.   \citet{2011ApJS..197...33S} obtained a larger size using a King profile with fixed concentration. The \citet{2013A&A...558A..14M} effective radius is the average of  the \citet{2005ApJ...627..203H} radii in the two filter bands, just as we use the average of the radii from the two deep images.

\citet{2005ApJ...627..203H} calculated the dynamical mass from their King model fit to the ACS VCS surface brightness profile, using the central velocity dispersion, which they obtained from the expected velocity dispersion profile by taking into account the seeing. 
\citet{2013A&A...558A..14M} corrected the calculation of  \citeauthor{2005ApJ...627..203H} to the proper King core radius and obtained a slightly larger dynamical mass estimate.
Our  dynamical mass for S999 is larger than both these values, but  in agreement  within the uncertainties with the corrected mass estimate by \citet{2013A&A...558A..14M} (see Table~\ref{table:S999lit}).

The magnitudes show an excellent agreement with \citet{2005ApJ...627..203H}, better than our estimate of their uncertainty. Their contribution to the uncertainty of the mass ratio is smaller than that of any of the other parameters.

\citet{2005ApJ...627..203H} did not calculate the mass ratio and quoted the dynamical mass-to-light ratio instead. They listed two metallicity estimates for S999, based on the ACS VCS $g-z$ colour and a $C-T_1$ colour, respectively. While the former is lower than our measurement, the latter  is reasonably close.
For calculating the stellar mass, \citet{2013A&A...558A..14M} assumed an age of 13 Gyr and extrapolated the \citet{2005MNRAS.362..799M} stellar mass-to-light ratios for S999's metallicity (they used $[Fe/H]=-1.4$) to get $M/L_V=2.2$.
\citet{2014MNRAS.444.2993F} used the stellar mass of  \citet{2014MNRAS.443.1151N}, which  used $M/L_V=1.1$ (or effectively, with S999's metallicity, an age of about 3.5 Gyr).
While we chose to use the $z$-band instead of the $V$-band,  the mass-to-light ratio in the $V$-band for the stellar populations we find is $M/L_V \sim 1.7$.

As we have seen, the dynamical mass estimates for S999 span a range of values, but agree within the uncertainties.
The other main driver for the large range of values for the dynamical-to-stellar mass ratio (a factor of $\sim$2 between the lowest and highest estimate) is the characterisation of the stellar populations. Our stellar mass-to-light ratio lies in-between the estimates of \citet{2013A&A...558A..14M} and \citet{2014MNRAS.444.2993F}. Also our preferred mass ratio is in-between the extremes of these two studies. However, in comparison to those, we clearly improved the characterisation of the stellar populations with the  ESI spectrum. Also with the better constrained stellar mass the dynamical-to-stellar mass ratio of S999 remains elevated, i.e.~with the preferred value of $M_{dyn}/M_*=8.2$ and a range from 5.6 to 11.2.

\subsection{Elevated mass ratios in UCDs}
Our mass ratio of $\sim$8 implies either  an underestimated stellar mass or extra mass that contributes to the dynamical mass without adding light.
In the following we discuss possible explanations for an elevated mass ratio in S999 and in UCDs in general.

\subsubsection{Non-canonical initial mass function}
If the initial mass function (IMF)  differs from the assumed one, then the inferred stellar mass changes, mostly increases.
This is true for both top and bottom-heavy IMFs, since  stellar remnants of massive stars add mass but no light, and low-mass stars contribute mass, while adding very little light. 
Both cases have been suggested to explain the elevated mass ratios in UCDs
\citep[e.g.][]{2008MNRAS.386..864D,2008AN....329..964M}.

However, by changing the assumed IMF, the fitted SSP age and metallicity will also be altered. 
In order to estimate the combined effect on the mass-to-light ratio we fitted \textsc{miles} SSP models with varying IMF slope
\citep[with slopes from 0.3 to 3.3, solar $\alpha$-abundances]{vazdekis2010}. 
With increasing slope, the fitted age generally also increases (as well as $\chi^2$), while the metallicity  decreases. 
 The \citeauthor{vazdekis2010} models are provided for IMFs with a power-law and a broken
power-law (with the slope varying for stars with $M>0.6 M_\odot$, see Appendix A of \citealt{vazdekis2003} for details).
Relative to the assumption of a Kroupa IMF the mass-to-light ratio increases by about a factor of three for  the 
broken power-law IMF with the steepest slope for high-mass stars  (and by about a factor of four for the shallowest).
Using a  single power-law IMF, the mass-to-light increases by a factor of about  3  for a slope of 2.3 as compared to 
a Salpeter IMF  (slope 1.3 in \citeauthor{vazdekis2010}'s parametrisation, with an increased mass-to-light ratio by about a third as compared to the Kroupa IMF; the shallowest single power-law IMF with 
a slope of 0.3 results in an increase by a factor 2 to 3). 
Formally, a single power-law IMF with a  slope of around 3.3 would result in a mass-to-light ratio
sufficient to account for the dynamical mass. However, again formally, the models with shallower slopes 
should be preferred, since the models with slopes steeper than 2.3 let the $\chi^2$ increase dramatically (by a factor of more than $>24$ for a slope of 3.3).\footnote{This most extreme model reverts the trend of fitted age increasing with the IMF slope, and results in the youngest fitted age, possibly as well indicating that the fit
is not reliable.} We note that such a slope is also beyond any observed IMF slope for the most massive galaxies \citep[e.g.][]{cappellari12,spiniello12,cvd12}.
Assuming that similar factors for the mass-to-light ratios would result from varying the IMF in the \citet{2011MNRAS.412.2183T} models, 
which we used for our mass-to-light estimation in order to account for the $\alpha$-enhancement,  
we conclude that none of the reasonable fits yields a mass-to-light ratio sufficient to result in a dynamical-to-stellar mass ratio of unity.

We note that \citet{2014AN....335..486F} did not find any evidence for a bottom-heavy IMF in S999 (nor S928), using the CO index. However, their uncertainties were too large to completely rule out a bottom-heavy IMF.
Recently, \citet{2012ApJ...747...72D} claimed that UCDs have high rates of X-ray binaries, indicative of
of an overpopulation of remnants of massive stars (however, \citealt{2013MNRAS.433.1444P} find the opposite for Fornax UCDs). 
We do not find any of the X-ray sources of \citet[with an X-ray completeness limit of $L_X\sim5\times10^{39}$ erg s$^{-1}$]{Jordan:2004hza} to coincide with the position of any of our 8 UCDs. However, this does not put firm constraints on the number of low-mass X-ray binaries and the remnants of massive stars.
Thus, we conclude that the elevated mass ratio of S999 is unlikely explained with a bottom-heavy IMF, but may have a contribution from an excessive number of stellar remnants from a top-heavy IMF.

\subsubsection{Central massive black hole}
\citet{2013A&A...558A..14M} showed that on average a massive black hole with a mass of about 15\% of the UCD stellar mass could explain the generally elevated mass ratios seen in UCDs, and is consistent with  progenitor galaxies having a total mass of about $10^9M_{\odot}$. 
For S999 they calculated the mass of the central black hole to be $M_{BH} = 25 \times 10^6 M_{\odot}$, several times the stellar mass, in order to explain  their mass ratio for S999. 
This would make S999  the most black hole dominated UCD or galaxy known by far. 

\citet{2014Natur.513..398S} have confirmed a massive black hole in M60-UCD1 (note that M60-UCD1's mass ratio was not previously found to be elevated based on its unresolved velocity dispersion).
This kind of data, however, is currently rare. \citet{2011MNRAS.414L..70F} did not find evidence for a massive black hole in  UCD3  (a UCD in the Fornax cluster which does not have an elevated mass ratio). The non-detection of S999 in the X-rays cannot rule out a massive black hole, but requires that it is non-accreting.
Integral field spectroscopy with high spatial resolution will allow one to detect the central black hole dynamically, or to put limits on its mass.

\subsubsection{Tidal Stripping}
\citet{2014MNRAS.444.2993F}  argued with  the stripping simulations of 
\citet{2013MNRAS.433.1997P}  that the early stages of the tidal stripping process may be an explanation for the observed extreme mass ratios seen in half a dozen UCDs, including S999. The mass ratios in the stripping simulations are reduced when taking into account the variation of velocity dispersion and virial coefficient (Pfeffer, priv.\ comm.), instead of assuming them to be constant like \citet{2014MNRAS.444.2993F} did. While they remain elevated, this puts even stronger constraints on the specific orbits that can cause stripping to be strong enough to explain the extreme mass ratios. 
If stripping is the cause of the elevated mass ratio we would expect the UCDs to reveal extra halo light in their surface brightness profiles. Our results presented here indicate that S999 has little if any extra halo light beyond a single S\'ersic profile. 
Thus we conclude it is unlikely that the extreme mass ratio of S999 is completely due to it currently undergoing tidal stripping. 
Several other UCDs with two-component structures  (VUCD7, UCD3, UCD5, FCC303, M60-UCD1, M59cO; \citealt{2005ApJ...627..203H,2003Natur.423..519D,2008MNRAS.385L..83C,2008AJ....136..461E,2011ApJ...737...86C}), possibly including two of the UCDs in our sample (S928 and S8005, see Appendix~\ref{sec:profiles}), 
are more likely candidates, and may suggest a continuity from nucleated galaxies via UCDs with envelopes to single component UCDs.

However, stripping is still part of the explanation, if a  central massive blackhole inherited from the more massive progenitor galaxy which had been stripped, is responsible for the elevated mass ratio.
Also, the positive colour gradients found for S999 and some of the other analysed UCDs might be expected from the stripping scenario:
the assumed progenitors, which were nuclei of early-type dwarf galaxies, typically have bluer colours than the host galaxy  \citep{2004ApJ...613..262L,2006ApJS..165...57C,Turner:2012fb,2011MNRAS.414.3052D}.
Generally, early-type galaxies with masses of the potential progenitors do not show negative colour gradients (as seen in more massive galaxies) but often have positive gradients  \citep{2010MNRAS.407..144T}, generally interpreted as age gradients.
The remnants in the stripping simulations of 
\citet{2013MNRAS.433.1997P} are larger in size than  the nuclei of the progenitor galaxies.
The authors  did not attribute this increase in size to dynamical heating of the nucleus,
  but rather to stars from the progenitor remaining in the remnant. 
  As these stars are expected to be typically found at larger radii than the stars of the nucleus,
  the observed positive colour gradient might be expected. 
The typical difference  in colour between the nucleus and the main body of the galaxy \citep[e.g.][]{2006ApJS..165...57C,2011MNRAS.414.3052D} is similar 
to the typical observed colour gradient in UCDs.

While \citet{2008AJ....136..461E} argued that the levels of $\alpha$-enhancements in UCDs and nuclei are not consistent with the stripping scenario,  \citet{2010ApJ...724L..64P} found them to be similar when comparing nuclei and UCDs at similar local projected galaxy density. 
The level of $\alpha$-enhancement in S999 is similar to that of \citeauthor{2010ApJ...724L..64P}'s high local density UCDs.

\subsubsection{Dark Matter}
\citet{2005ApJ...627..203H} suggested (non-baryonic) dark matter as a possible explanation for the elevated mass ratio of S999.
 However,  \citet{2009ApJ...691..946M} argued that, by comparison to the central dark matter densities of theoretical dark matter profiles and observed Local Group dwarf spheroidals, dark matter cannot contribute large amounts of mass in UCDs. Typically, the dark matter fraction within the half-light radius is expected to be less than 30\% \citep[see also][]{2011ApJ...726..108T,2014Natur.513..398S}.

In their stripping simulations, \citet{2013MNRAS.433.1997P}  assumed that the progenitor galaxy is largely stripped of dark matter before the stars can be affected, and did not include any dark matter. 
 \citet{2008MNRAS.391..942B} showed that mass segregation in compact stellar systems will expel dark matter particles from the dense central regions of these objects. However, the authors calculated that the time scale for this process is larger than a Hubble time for UCDs.

 We conclude that, while S999 is a good candidate to search for dark matter in UCDs thanks to its low stellar density compared to other UCDs, dark matter should not play a major
 role in explaining the elevated mass ratio.

\section{Summary}

We used our measurements of size ($R_e=20.9\pm1.0$ pc) and velocity dispersion ($\sigma_{ap}=27\pm2$ km s$^{-1}$)
based on high-quality data to determine the dynamical mass ($M_{dyn}=32\pm5 \times 10^6 M_{\odot}$) of the ultra-compact dwarf (UCD) S999 via mass modelling.
Furthermore, we newly obtained stellar population characteristics 
(Age $=7.6^{+2.0}_{-1.6}$ Gyr; $[Z/\textrm{H}]=-0.95^{+0.12}_{-0.10}$;  $[\alpha/\textrm{Fe}]=0.34^{+0.10}_{-0.12}$),
and calculated the stellar mass $M_*=3.9^{+0.9}_{-0.6}\times10^6 M_{\odot}$ assuming a Kroupa stellar initial mass function (IMF).
With these new independent measurements we  derived the dynamical-to-stellar mass ratio of S999 $M_{dyn}/M_*=8.2$.
Furthermore, we  estimated realistic uncertainties of the measurements, and considered the possible ranges of the values.
We confirmed the  mass ratio to be elevated beyond what is expected from stellar populations with a canonical IMF for the whole range of possible values $5.6 \le M_{dyn}/M_* \le 11.2$.

This elevated mass ratio can potentially offer insights to the formation of S999 and other UCDs, and we discussed possible, non-exclusive explanations. New measurements, e.g.~of  IMF-sensitive features of the spectrum and high-resolution integral field spectroscopy, are needed to explore the contributions of the listed possibilities.

We conclude that  the most likely options include a central massive black hole (as recently found in M60-UCD1),  and  that the object is not in equilibrium so that the mass estimate based on the velocity dispersion is not adequate.
Both might be related to S999 being the remnant nucleus of a tidally stripped progenitor galaxy, which could also explain its positive colour gradient.
Most likely a combination of several of the explanations is required to explain the extreme value of the mass ratio of S999, since none of the possibilities alone is likely to  account for all of the excess dynamical mass compared to the mass in stars.

\section*{Acknowledgments}
The authors thank the referee for helpful comments.
They also thank M.~Onodera for providing the code for interpolating the stellar population models, and T.~Lisker for the non-parametric photometry code.
JJ also thanks M.T.~Murphy and J.~Pfeffer for useful discussions about reduction of spectra and stripping simulation, respectively. 
JJ and DAF thank the
ARC for financial support via DP130100388.
Based on observations obtained with WIRCam, a joint project of CFHT, Taiwan, Korea, Canada, France, at the Canada-France-Hawaii Telescope (CFHT) which is operated by the National Research Council (NRC) of Canada, the Institut National des Sciences de l'Univers of the Centre National de la Recherche Scientifique of France, and the University of Hawaii.

\appendix
\section{Light profiles of UCDs close to M87}
 \label{sec:profiles}
 In addition to S999 there are seven more UCDs in the ACS field of view. Here we present the surface brightness and colour profiles of the remaining UCDs.
Most of the analysed UCDs follow single S\'ersic profiles well (Fig.~\ref{fig:profiles}).
While the proper subtraction of M87's light is non-trivial, the consistency of
the profiles and their fits in the two filter bands is reassuring.
Two UCDs, S928 and S8005, show a clear dip of the observed profile in comparison
to the fitted profile (seen as a positive residual) at a surface brightness around 20 to 21 mag arcsec$^{-2}$ in the $I$ band.

The colour gradients in the UCDs are mostly positive. Again, the uncertainty of the background subtraction is the major limitation. However, the colour profiles are consistent with the $g-z$ profiles measured on the ACS VCS images. For at least three UCDs -- S999, S887, and S5065 -- the colours  become steadily consistently bluer by about $\sim$ 0.15 mag towards larger radii over a range where the colours are reliable.
The fourth cosine term profile, i.e.~disky or boxy isophotes, does  not generally display significant features  correlated with the surface brightness profile -- with  the possible exception of S8005, which becomes disky in the outer parts ($C_4 \sim 0.04$).

Our photometric measurements of the UCDs in the ACS VCS pointing are given in Table \ref{table:ACSphotometry}.

\begin{table*}
\caption{Imaging of UCDs from the ACS VCS}
\begin{center}
\begin{tabular}{l|cc|cc|cccc|cccc}
\hline
UCD & \multicolumn{4}{c}{Non-parametric}  & \multicolumn{8}{c}{S\'ersic fit} \\
 & \multicolumn{2}{c}{F475W ($g$)}  & \multicolumn{2}{c}{F850LP ($z$)}  & \multicolumn{4}{c}{F475W ($g$)}  & \multicolumn{4}{c}{F850LP ($z$)} \\
 & $m$ & $a_e$ [$\arcsec$] & $m$ & $a_e$ [$\arcsec$]  & $m$ & $a_e$ [$\arcsec$] & $\mu_e$ & $n$ & $m$ & $a_e$ [$\arcsec$] & $\mu_e$ & $n$ \\
\hline
    S999 &  20.38 &  0.28 & 19.49 &  0.29  &  20.31 &  0.26 & 20.04&    1.1  & 19.42 &  0.28 & 19.33 &   1.0 \\
  H44905 &  22.07 &  0.26 & 21.09 &  0.27  &  22.00 &  0.26 & 21.62&    0.7  & 21.12 &  0.29 & 20.82 &   0.5 \\
    S887 &  21.29 &  0.17 & 20.23 &  0.19  &  20.92 &  0.13 & 19.41&    2.0  & 19.93 &  0.16 & 18.94 &   2.1 \\
    S928 &  19.97 &  0.30 & 19.13 &  0.30  &  19.82 &  0.29 & 19.84&    1.3  & 18.95 &  0.31 & 19.11 &   1.2 \\
  S5056 &  20.31 &  0.20 & 19.35 &  0.22  &  20.14 &  0.18 & 19.14&    1.0  & 19.10 &  0.19 & 18.22 &   1.2 \\
   S8005 &  20.71 &  0.32 & 19.84 &  0.33  &  20.59 &  0.33 & 20.75&    1.1  & 19.62 &  0.35 & 19.98 &   1.3 \\
   S8006 &  20.67 &  0.25 & 19.78 &  0.26  &  20.56 &  0.24 & 20.13&    1.1  & 19.57 &  0.25 & 19.27 &   1.2 \\
\hline\end{tabular}
\end{center}
Same as Table \ref{table:photometry}, but for the ACS VCS images in the $g$ and $z$-bands. S672 is outside the field of view.
\label{table:ACSphotometry}
\end{table*}

\begin{figure*}
\includegraphics[height=0.9\textwidth,angle=-90]{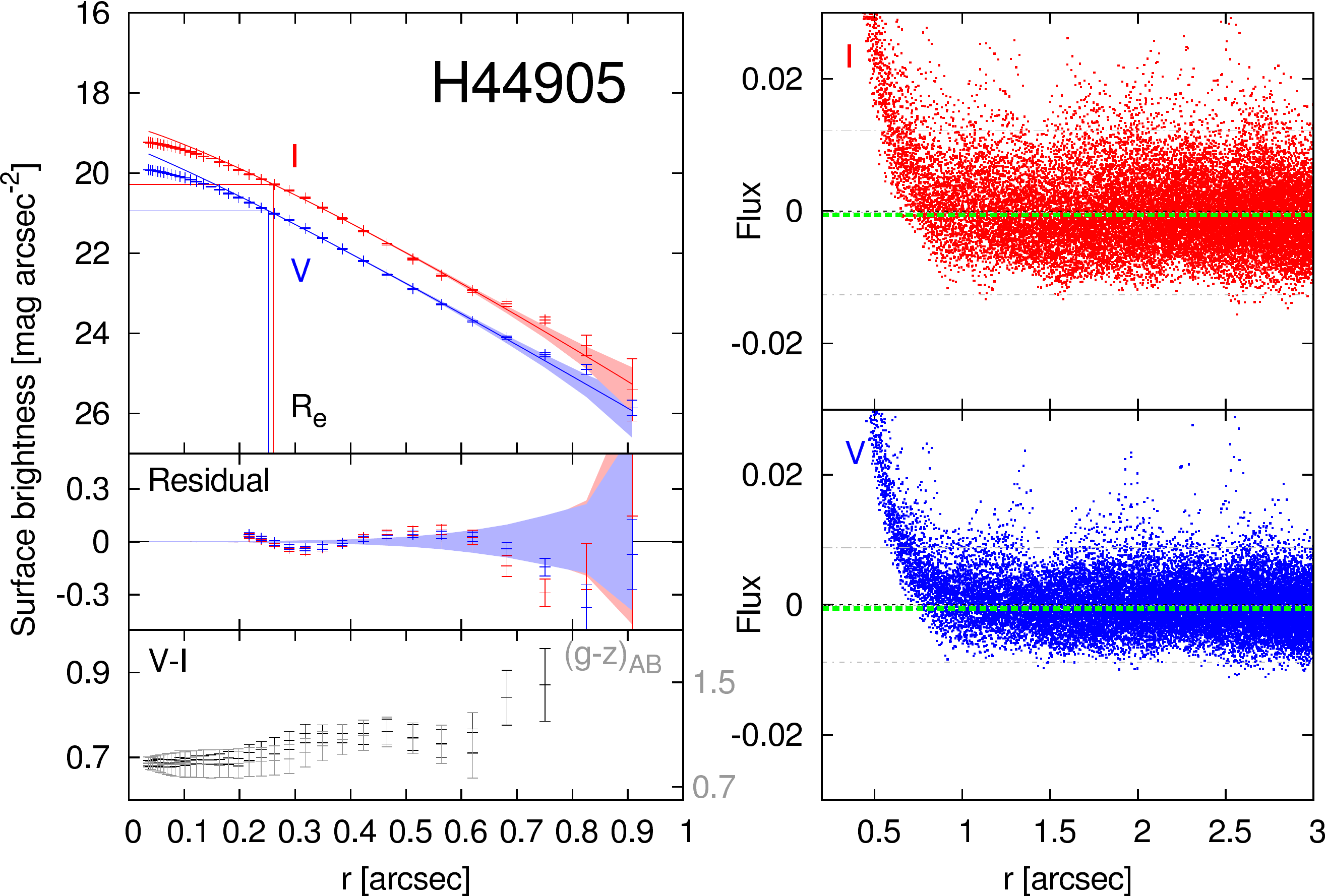}\\~\\

\caption{Surface brightness and colour profiles  of UCDs around M87. Same as Fig.~\ref{fig:profile}. The name of the UCD is given in the top left panel. \label{fig:profiles}}
\end{figure*}

\begin{figure*}
\contcaption{}
\includegraphics[height=0.9\textwidth,angle=-90]{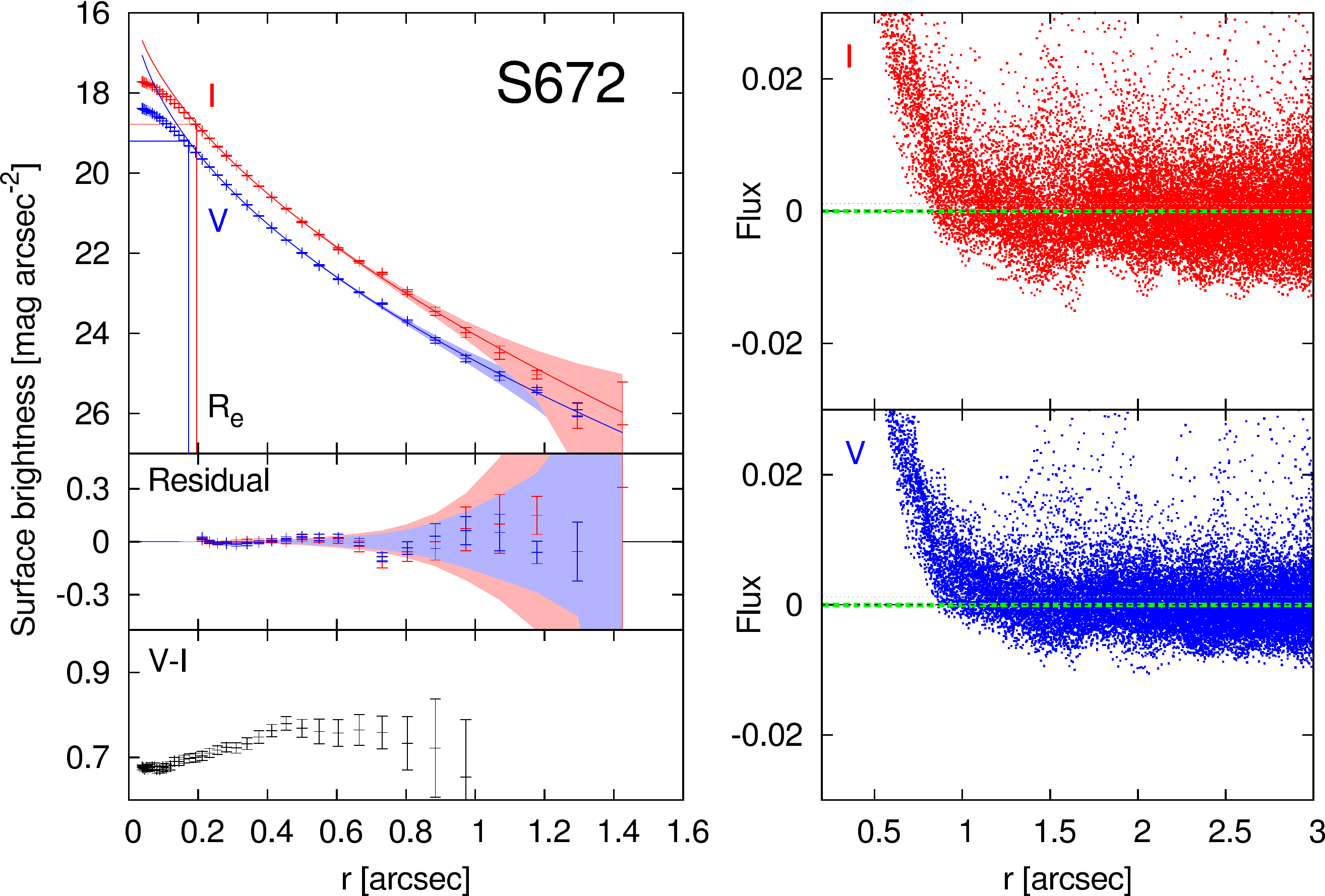}\\~\\
\includegraphics[height=0.9\textwidth,angle=-90]{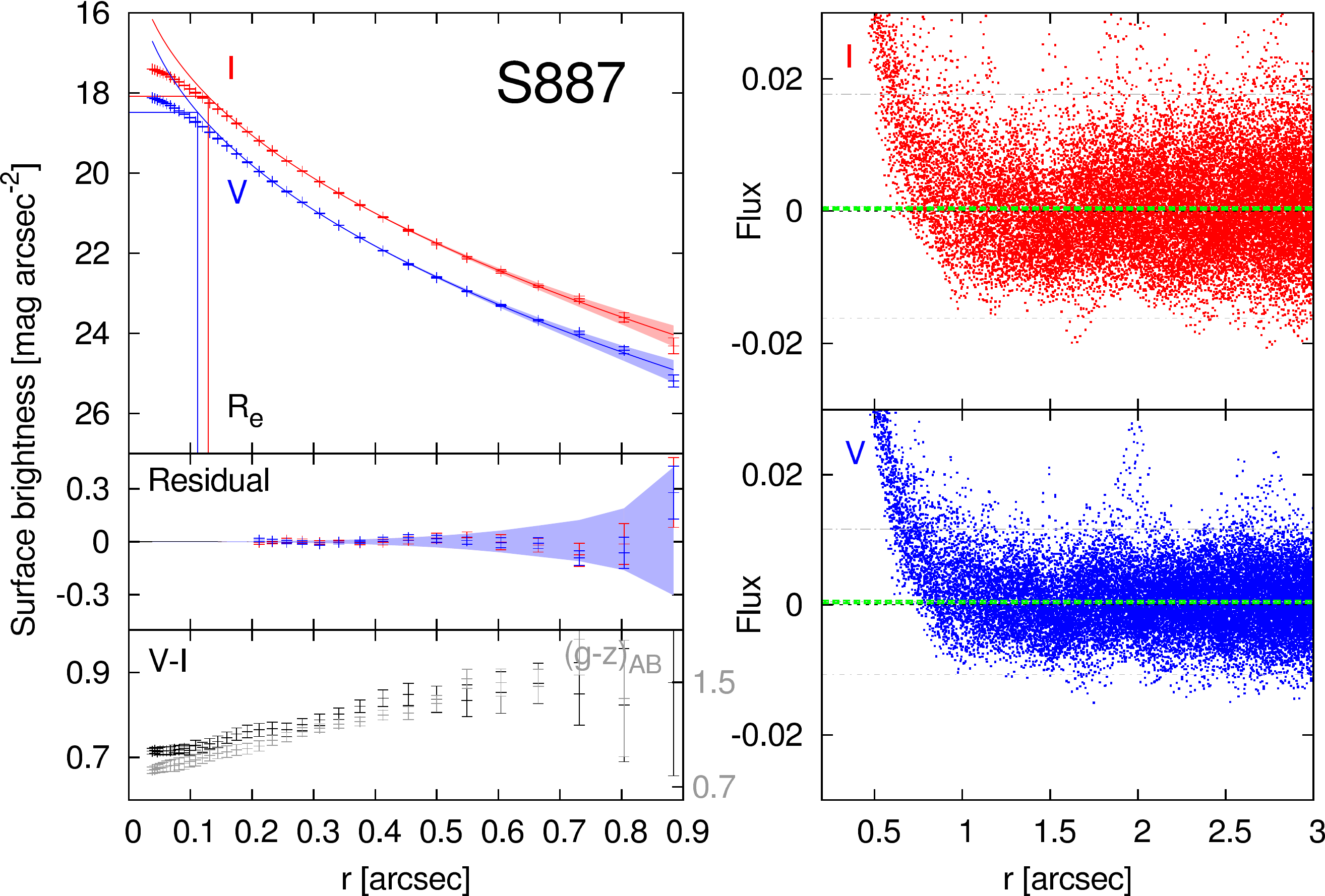}
\end{figure*}

\begin{figure*}
\contcaption{}
\includegraphics[height=0.9\textwidth,angle=-90]{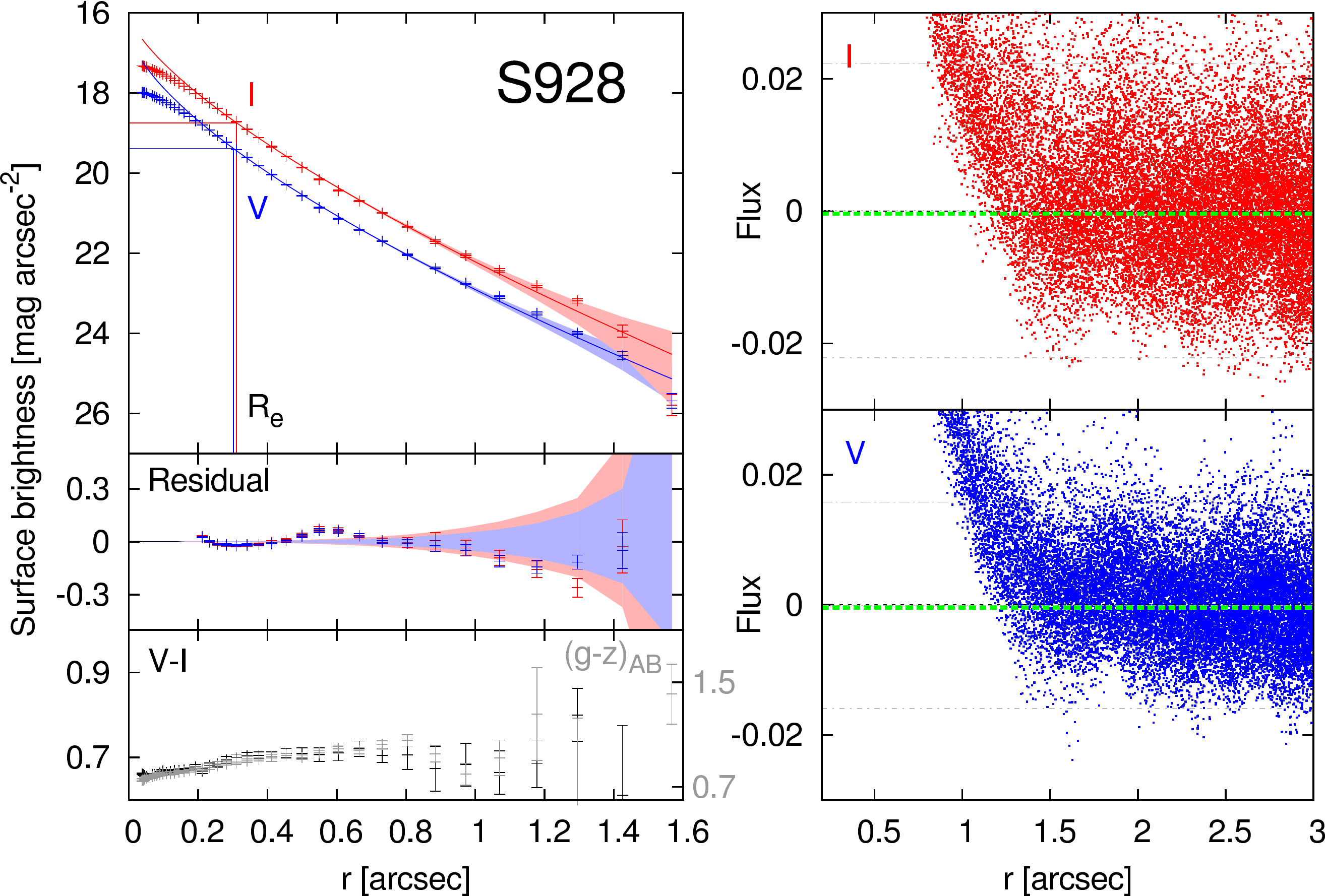}\\~\\
\includegraphics[height=0.9\textwidth,angle=-90]{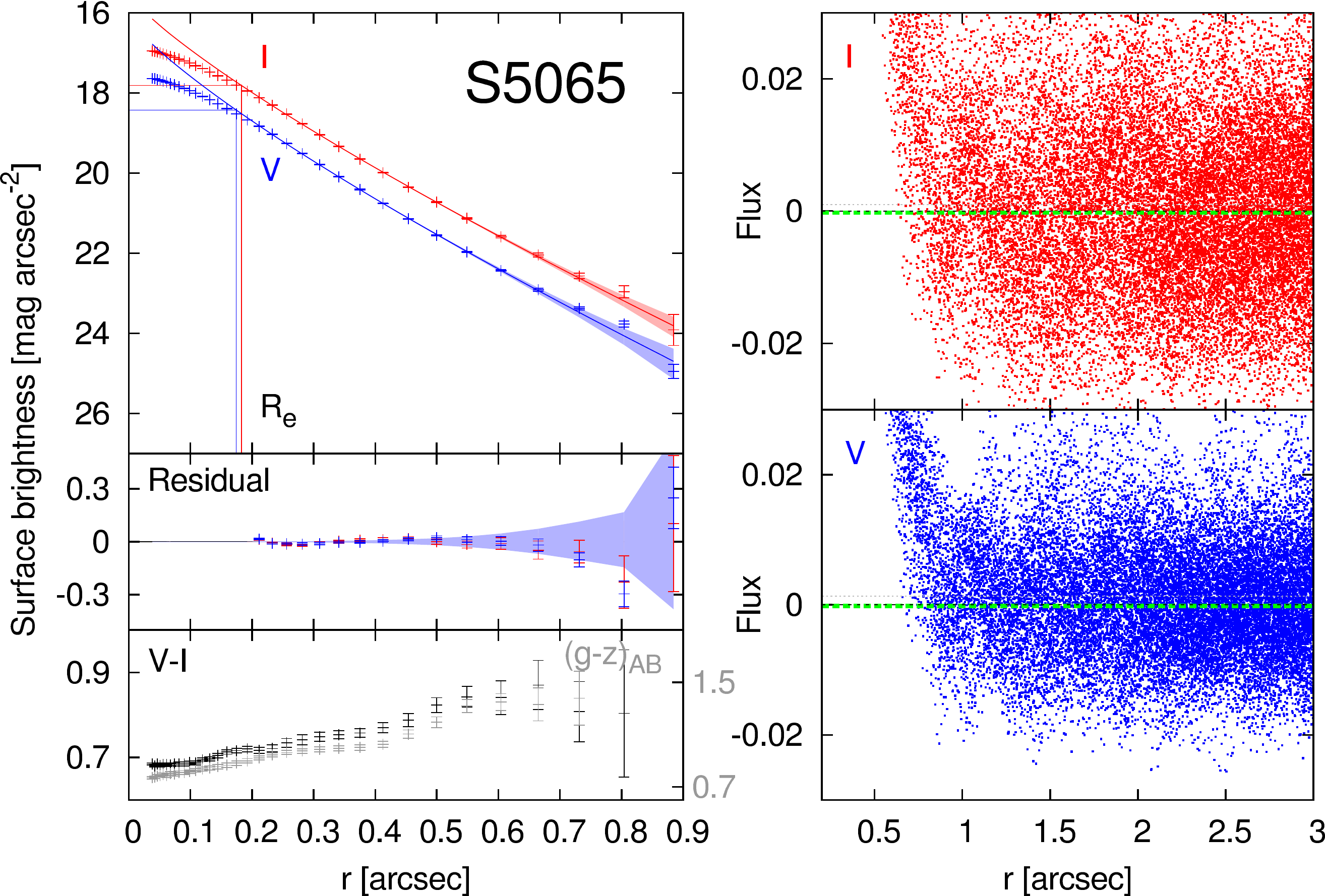}
\end{figure*}

\begin{figure*}
\contcaption{}
\includegraphics[height=0.9\textwidth,angle=-90]{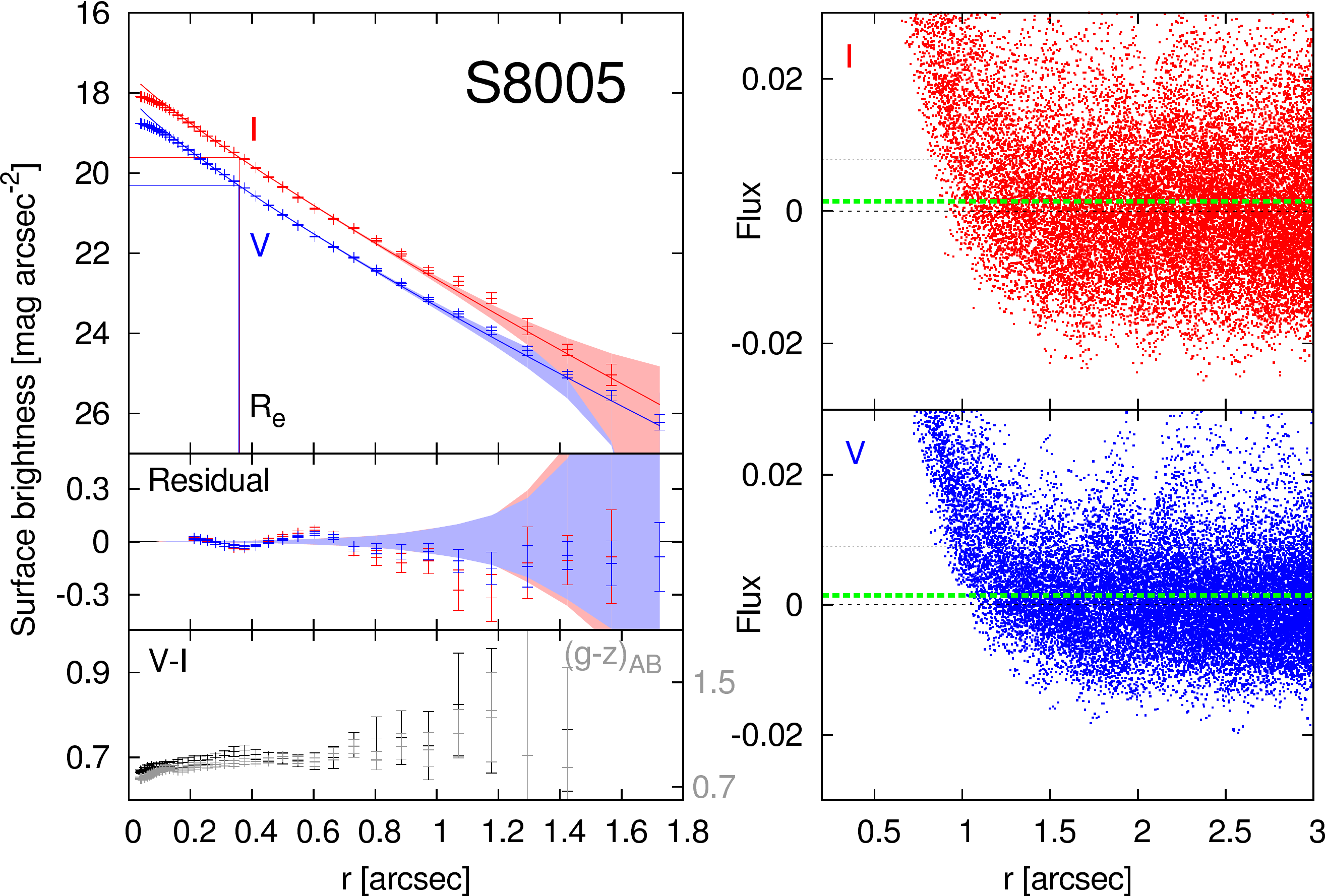}\\~\\
\includegraphics[height=0.9\textwidth,angle=-90]{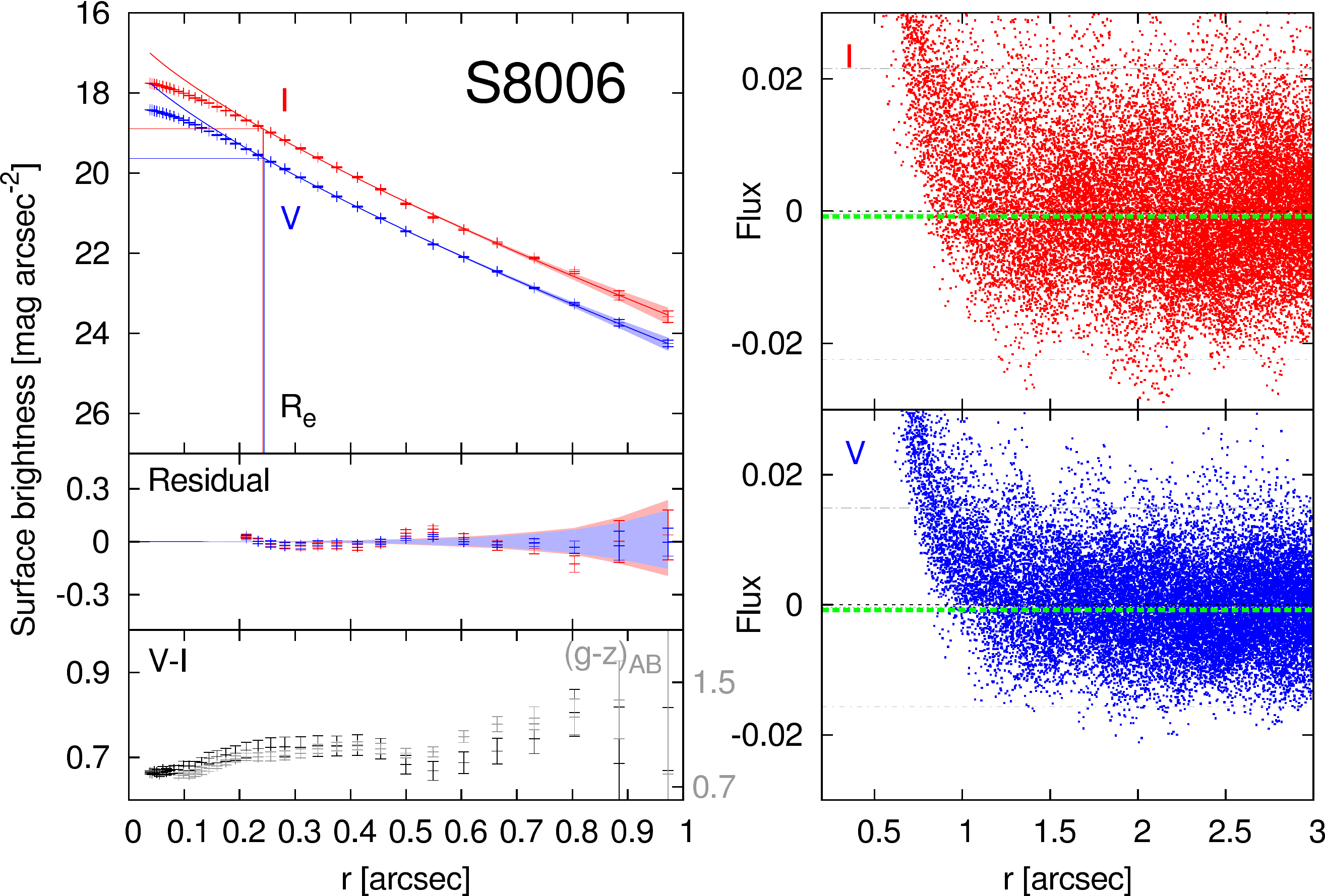}
\label{lastpage}
\end{figure*}

\end{document}